\documentclass[prd,aps,twocolumn,nofootinbib,showpacs,superscriptaddress]{revtex4-1}
\usepackage{amsfonts}
\usepackage{amsmath}
\usepackage{amssymb}
\usepackage{bm}
\usepackage{dcolumn}
\usepackage{graphicx}
\usepackage{graphics}
\usepackage[latin1]{inputenc}
\usepackage{latexsym}
\usepackage{rotating}
\usepackage{hyperref}
\usepackage{xspace} 
\usepackage[usenames]{color}
\usepackage{mathrsfs}
\usepackage{multirow}

\usepackage{longtable}
\usepackage{pict2e}
\usepackage[usenames,dvipsnames]{xcolor}
\usepackage{url}
\usepackage{gensymb}

\allowdisplaybreaks

\definecolor {darkgreen}{rgb}{0.2,0.7,0.2}
\definecolor{purple}{rgb}{0.5,0,0.5}

\newcommand{\bea}{\begin{eqnarray}}
\newcommand{\eea}{\end{eqnarray}}
\newcommand{\beq}{\begin{equation}}
\newcommand{\eeq}{\end{equation}}

\newcommand{\harm}{{\sc Harm3d}\xspace}

\def\fun#1#2{\lower3.6pt\vbox{\baselineskip0pt\lineskip.9pt
  \ialign{$\mathsurround=0pt#1\hfil##\hfil$\crcr#2\crcr\sim\crcr}}}

\def\lambdabar{%
\relax
\bgroup
\def\@tempa{\hbox{\raise.73\ht0
\hbox to0pt{\kern.25\wd0\vrule width.5\wd0
height.1pt depth.1pt\hss}\box0}}%
\mathchoice{\setbox0\hbox{$\displaystyle\lambda$}\@tempa}%
{\setbox0\hbox{$\textstyle\lambda$}\@tempa}%
{\setbox0\hbox{$\scriptstyle\lambda$}\@tempa}%
{\setbox0\hbox{$\scriptscriptstyle\lambda$}\@tempa}%
\egroup
}

\begin{document}

\title{Approximate black hole binary spacetime via asymptotic matching}

\author{Bruno C. Mundim}
\affiliation{Center for Computational Relativity and Gravitation,
and School of Mathematical Sciences, Rochester Institute of
Technology, Rochester, New York 14623, USA.}
\affiliation{Max-Planck-Institut f{\"u}r Gravitationsphysik,
Albert-Einstein-Institut, D-14476 Golm, Germany.}

\author{Hiroyuki Nakano}
\affiliation{Center for Computational Relativity and Gravitation,
and School of Mathematical Sciences, Rochester Institute of
Technology, Rochester, New York 14623, USA.}
\affiliation{Yukawa Institute for Theoretical Physics, Kyoto University,
Kyoto 606-8502, Japan.}

\author{Nicol\'{a}s Yunes}
\affiliation{Department of Physics, Montana State University,
Bozeman, Montana 59717, USA.}

\author{Manuela Campanelli}
\affiliation{Center for Computational Relativity and Gravitation,
and School of Mathematical Sciences, Rochester Institute of
Technology, Rochester, New York 14623, USA.}

\author{Scott C. Noble}
\affiliation{Center for Computational Relativity and Gravitation,
and School of Mathematical Sciences, Rochester Institute of
Technology, Rochester, New York 14623, USA.}

\author{Yosef Zlochower}
\affiliation{Center for Computational Relativity and Gravitation,
and School of Mathematical Sciences, Rochester Institute of
Technology, Rochester, New York 14623, USA.}

\begin{abstract}

We construct a fully analytic, general relativistic, nonspinning black hole binary spacetime that approximately solves
the vacuum Einstein equations everywhere in space and time for black holes sufficiently well separated.
The metric is constructed by asymptotically matching perturbed Schwarzschild metrics near each black hole
to a two-body post-Newtonian metric far from them, and a two-body post-Minkowskian metric farther still.
Asymptotic matching is done without linearizing about a particular time slice, and thus it is valid dynamically
and for all times, provided the binary is sufficiently well separated. 
This approximate global metric can be used for long dynamical evolutions of relativistic magnetohydrodynamical,
circumbinary disks around inspiraling supermassive black holes to study a variety of phenomena.

\end{abstract}

\pacs{04.25.Nx, 04.25.dg, 04.70.Bw} 
\maketitle

\section{INTRODUCTION}

The interaction between black holes and matter in the highly energetic and
strong gravitational regime can reveal invaluable information both about the
geometry of these dark objects, as well as the physics of magnetohydrodynamics.
A particularly interesting astrophysical scenario is a circumbinary accretion
disk that surrounds a supermassive black hole (SMBH) binary. As the SMBHs
slowly spiral toward each other, plunge and eventually merge, the interactions
between gravity and matter lead to the emission of strong electromagnetic (EM)
radiation and the formation of jets. Instruments such as
Pan-STARRS~\cite{2010SPIE.7733E..12K} or the planned
LSST~\cite{2009arXiv0912.0201L}, are designed to detect such events and
characterize them up to cosmological distances (see~\cite{Schnittman:2013qxa} for a recent review on this subject).

The search and characterization of such energetic events can be aided by
predicting their EM features, but such predictions require modeling. The
modeling of an accretion disk about an inspiraling, SMBH binary is a
challenging problem because one needs to solve the general relativistic
magnetohydrodynamic (GRMHD) equations---to evolve the disk, resolve shocks, and
instabilities---coupled to the full Einstein equations---to evolve the SMBH
binary. In spite of the many recent breakthroughs in numerical
relativity~\cite{Pretorius:2005gq, Campanelli:2005dd, Baker:2005vv}, a full
numerical solution to the Einstein-GRMHD coupled system over many orbits is far
too computationally expensive (see e.g.,~\cite{Farris:2011vx, Bode12,
Giacomazzo:2012iv}). 

The modeling of circumbinary accretion disks
would be greatly aided if one could calculate the SMBH binary spacetime
fully analytically.  The freedom to discretize the spacetime domain 
most appropriately for the matter evolution without also having to maintain
a stable evolution of the Einstein equations leads to a more efficient and 
accurate simulation of the accretion disk. This is because the 
Courant condition greatly limits the time step size,
making full numerical simulations of the spacetime 
impractical when the characteristic MHD speeds are significantly 
smaller than $c$.

The modeling of compact binaries is well suited to the post-Newtonian (PN)
approximation (see e.g.~\cite{Blanchet:2006zz}), 
where one solves the Einstein equations in a weak-field and a slow-motion expansion. 
The latter is an excellent approximation to the inspiral of compact objects, 
since their orbital velocity is much smaller than the speed of light until right before they plunge.
The former, however, is a poor approximation to describe black holes, as their gravitational field 
is not weak close to the event horizon. But it is precisely
this region that is of most astrophysical interest---where jets 
are launched, matter is swallowed up by the SMBHs, and individual
accretion disks may form.

In spite of this, a PN spacetime was used in~\cite{Noble:2012xz} to study circumbinary disks, 
using the \harm code~\cite{Noble09} to solve the GRMHD equations. Because the PN approximation breaks down close
to event horizons, Ref.~\cite{Noble:2012xz} was forced to excise the 
region encompassing the binary, precisely where one may expect very interesting EM and fluid behavior.
Nonetheless, that study showed that a nontrivial and variable EM signal
originated from an overdense region close to the inner edge of the disk. This
{\it lump} arose after many tens of binary orbits, 
and therefore it had previously only been seen in Newtonian
simulations~\cite{Shi:2011us}. 

The main goal of this paper is to combine the PN approximation with other black hole
perturbation theory ideas~\cite{Yunes:2005nn} to construct a global, purely analytical
spacetime that is approximately valid \emph{everywhere} in space (including close to the event horizon)
and time (up until roughly $10 M$, with $M$ the total mass). 
This metric can then be used as a background on which to solve the GRMHD equations
to evolve magnetized circumbinary disks everywhere on the computational domain. 
In particular, it will allow the BHs to actually be on the numerical domain, making circumbinary excision unnecessary
and allowing for the modeling of all relevant physics close to each event horizon.
This includes the study of disk formation around individual BHs, the relativistic dynamics of gas
in the interior circumbinary region and the formation of jets and shocks.  

The global metric will be constructed by first subdividing the spacetime into three
separate regions, where different assumptions hold and different approximations
can be used. The {\emph{inner zone}} (IZ) is the region sufficiently close to
either black hole where the metric can be described by a perturbed Schwarzschild
or Kerr solution~\cite{Thorne:1980ru,
Thorne:1984mz, Detweiler:2005kq, Yunes:2005ve,Chatziioannou:2012gq}.  The {\emph{near zone}} (NZ) is
the region far away from either black hole, but less than a gravitational wave
wavelength $\lambda$ from the binary's center of mass, such that the metric can
be described by a PN expansion~\cite{Blanchet:1998vx}. The {\emph{far}} or
{\emph{wave zone}} (FZ) is the region outside a gravitational wave wavelength
from the binary's center of mass, where the metric can be described by a
multipolar post-Minkowskian (PM) expansion~\cite{Will:1996zj}. Unlike in a PN
expansion, a PM treatment properly accounts for gravitational wave retardation,
which is essential in the wave zone. Of course, such a subdivision of the
spacetime is only valid when the SMBHs are sufficiently well separated and
slowly inspiraling, breaking down at separation of roughly $10 M$. 

A global spacetime can then be built from the IZ, NZ and FZ metrics by
asymptotically matching them inside overlapping regions of validity, or
\emph{buffer zones} (BZs), where adjacent metrics are simultaneously valid. The
matching procedure returns a coordinate and parameter transformation relating
adjacent metrics, such that the latter asymptote to each other in the BZ. Asymptotic matching
in GR was carried out in~\cite{Alvi:1999cw,Pati:2000vt, Pati:2002ux, Alvi:2003pn, Yunes:2005nn, Yunes:2006iw,
Yunes:2006mx, JohnsonMcDaniel:2009dq}, but always restricting attention to a
particular spatial hypersurface. In this paper, we are interested in long,
dynamical evolutions, and we thus lift this restriction, allowing for
time-dependent, asymptotically matched transformations. Once the metrics have been
asymptotically matched, a global spacetime is constructed through transition
functions~\cite{Yunes:2006mx}, carefully designed to avoid introducing spurious
errors in the spacetime larger than those already contained in the individual
approximate metrics. 

\begin{figure}[htb]
\begin{center}
\includegraphics[width=0.45\textwidth,clip=true]{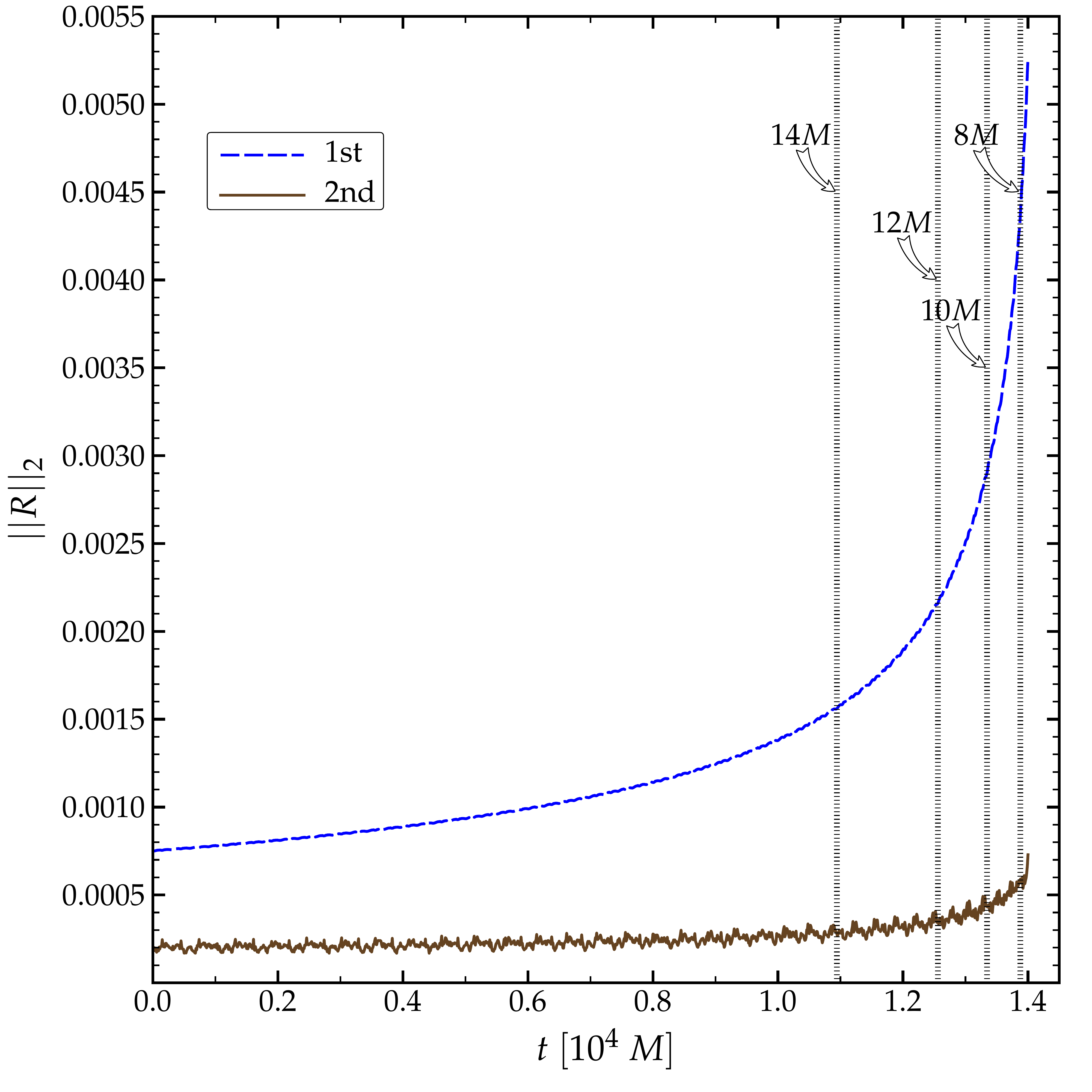}
\end{center}
\caption{(Color online) 
L2 norm of the Ricci scalar as a function of time computed with metrics
of two different perturbation orders (see Sec.~\ref{sec:global-metric}).  The
norms are computed using a physical domain extent of $[0,-80,-0.15]$ to
$[159.95,79.95,0.1]$ in units of total mass $M$. The black hole horizons are
excised and a mesh spacing of $0.05M$ is used.  The black, dotted, vertical lines indicate instants
of time when the binary separations reach $14M$,
$12M$, $10M$ and $8M$, from left to right, respectively.  The evolution is
initialized at $20M$. 
}\label{fig:l2_Ricci_new_coord}
\end{figure}

Such a global metric is an approximate solution to the vacuum Einstein
equations, satisfying the latter only to the degree that the individual
approximate metrics do. In order to determine the accuracy of the global
metric, we compute the Ricci scalar as a function of time, as shown in 
Fig.~\ref{fig:l2_Ricci_new_coord}, for two metric perturbation orders, which
will be defined in Sec.~\ref{sec:global-metric}. 
As expected the Ricci scalar is much smaller for the second-order metric, growing
with time as the inspiral proceeds and the orbital separation decreases.  
In this paper, we focus on nonspinning black hole binaries in a quasicircular
inspiral trajectories, but the methods introduced here are valid for general
black holes in generic orbits.  

The remainder of this paper deals with the details of this calculation. 
Section~\ref{sec:global-metric} describes how to construct the global metric in
more detail. 
Section~\ref{sec:ricci} shows the degree to which the global metric satisfies
the vacuum Einstein equations. 
Section~\ref{sec:conclusions} concludes and points to future research.  
We mainly follow the conventions of Misner, Thorne and
Wheeler~\cite{MTW}. In particular, we use greek letters $(\alpha, \beta,
\cdots)$ in index lists to denote spacetime indices, and latin letters $(i, j,
\cdots)$ to denote spatial indices. The metric is denoted $g_{\mu \nu}$ and it
has signature $(-,+,+,+)$. We use the geometric unit system, where $G=c=1$,
with the useful conversion factor $1 M_{\odot} = 1.477 \; {\rm{km}} = 4.926
\times 10^{-6} \; {\rm{s}}$. 

\section{CONSTRUCTION OF APPROXIMATE GLOBAL METRIC}
\label{sec:global-metric}

In this section, we describe the construction of an approximate, global
spacetime for a non-spinning binary black hole system in a quasicircular
trajectory, during the inspiral regime. We begin by subdividing the spacetime
into different zones, inside which different approximations will be used to
obtain approximate metrics. We then explain how to connect these approximate
metrics through time-dependent asymptotic matching, and conclude with a
description of the transition functions necessary to glue these asymptotically matched, 
approximate metrics together. 

\subsection{Subdividing spacetime}

\begin{figure}[t]
\begin{center}
\includegraphics[width=0.48\textwidth,clip=true]{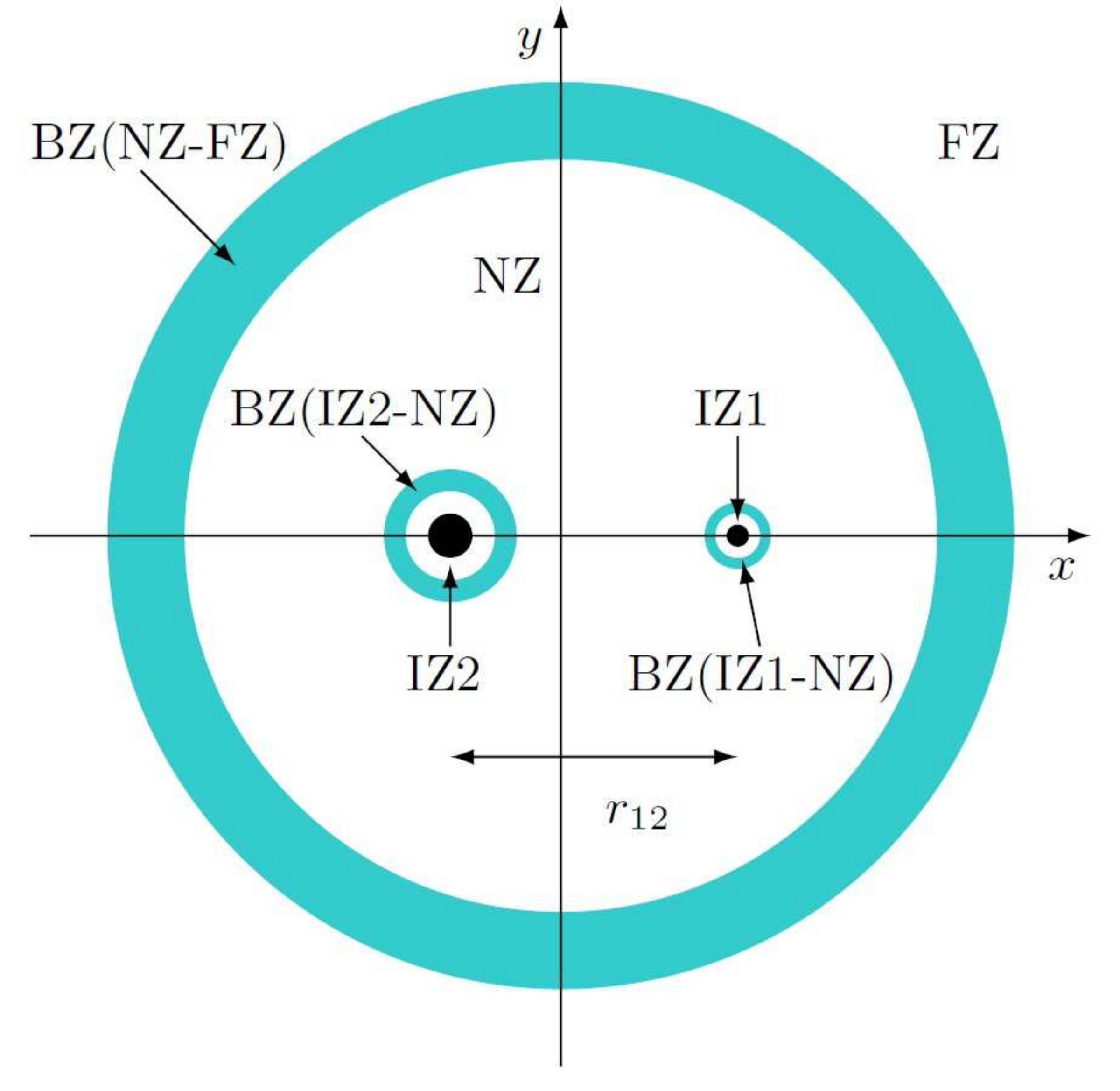}
\end{center}
\caption{(Color online)
Schematic diagram of the spacetime on a spatial hypersurface. BH1 and
BH2 are denoted by filled, solid black circles, where the orbital separation is
$r_{12}$. The BZs are denoted with cyan shells, the outer one representing the
FZ-NZ BZ and the two inner ones representing the NZ-IZ BZs (see also
Table~\ref{tab:zones}). The circular nature of these BZs is only
schematic; in practice, the BZs should be distorted. The IZ, NZ and FZ are also
shown in this figure.}
\label{fig:zones}
\end{figure}

We divide the spacetime into the following three regions:
\begin{itemize}
\item {\emph{Inner Zone}} for BH1 (IZ1) and BH2 (IZ2).
\item {\emph{Near Zone}} (NZ).
\item {\emph{Far Zone}} (FZ).
\end{itemize}
Figure~\ref{fig:zones} shows a schematic representation of these three zones on
a spatial hypersurface (see also~\cite{Gallouin:2012kb}). Black holes are shown
as black, solid circles, with cyan circles denoting the boundaries of each
zone, which are also listed in Table~\ref{tab:zones}.

IZ$A$ is defined as the region sufficiently near black hole $A$ that the metric
can be described as a tidally distorted black hole spacetime. Mathematically,
this region can be defined via $r_{A} \ll r_{12}$, where $r_{A}$ is the
distance from the $A$th black hole and $r_{12}$ is the binary's orbital
separation. The IZ metric can be modeled through black hole perturbation
theory, as in~\cite{Poisson:2005pi, Yunes:2005ve, Detweiler:2005kq,
Taylor:2008xy, Comeau:2009bz, Poisson:2009qj, Chatziioannou:2012gq}. In this
paper, we concentrate on nonspinning black hole binaries, and thus, the IZ
metrics will be described by a perturbed Schwarzschild solution either to quadrupole
or to octupole order~\cite{Poisson:2005pi,Detweiler:2005kq} in horizon-penetrating
coordinates~\cite{JohnsonMcDaniel:2009dq}. 

A highly desirable feature of the global metric for numerical purposes is the
use of {\it horizon penetrating}  Cook-Scheel harmonic
coordinates~\cite{Cook:1997qc}, which we will employ in the IZs. Such coordinates 
allow for excision of the region interior to the event horizons
(especially for matter falling into the black holes). Nonpenetrating
coordinates, such as standard PN harmonic coordinates, would lead to 
coordinate singularities and numerical difficulties
on the horizons. Asymptotic matching will not spoil the horizon-penetrating
properties of the global metric, since close to the horizons 
this metric will be that of the IZ in proper horizon-penetrating coordinates. 

\begin{table}[t]
\caption{Spatial regions of validity for the different zones.
$r_A$ is the distance from the $A$th black hole with mass
$m_A$, $r$ is the distance from the center of mass, and $r_{12}$ and $\lambda
\sim \pi \sqrt{r_{12}^3/(m_1+m_2)}$ are the orbital separation and the
gravitational wavelength, respectively. For BZs to exist, the system
must satisfy $m_A \ll r_{12}$.}
\label{tab:zones}
\begin{center}
\begin{tabular}{c|ccccc}
  Zone     & \multicolumn{5}{c}{Region of Validity}\\
  \hline\\
  IZ1      & $0     $&$  <    $&$  r_1  $&$  \ll  $&$ r_{12}$  \\
  IZ2      & $0     $&$  <    $&$  r_2  $&$  \ll  $&$ r_{12}$ \\
  NZ       & $m_A   $&$  \ll  $&$  r_A  $&$  \ll  $&$ \lambda$ \\
  FZ       & $r_{12} $&$  \ll  $&$  r    $&$  <    $&$ \infty$ \\
  IZ-NZ BZ & $m_A   $&$  \ll  $&$  r_A  $&$  \ll  $&$ r_{12}$  \\
  NZ-FZ BZ & $r_{12} $&$  \ll  $&$  r    $&$  \ll  $&$ \lambda$\\
  \end{tabular}
\end{center}
\end{table}

The NZ is the region sufficiently far from either black hole
that the metric can be described through the PN 
approximation~\cite{Blanchet:2006zz}, yet is still much closer
than a gravitational wavelength from the binary's center of mass.
Mathematically, this region can be defined via $m_{A} \ll r_A \ll \lambda$, where
$m_{A}$ is the mass of the $A$th black hole and $\lambda$ is the gravitational wave
wavelength.
In the PN approximation, one solves the Einstein equations as an
expansion in both $v/c \ll 1$ (slow motion), where $v$ is the binary's orbital velocity and
$c$ is the speed of light, and $Gm/(R c^2) \ll 1$ (weak fields), where $m$ is $m_{A}$ or
$M=m_A+m_B$, and $R$ is $r_{A}$ or $r_{12}$. By construction, the PN approximation
models black holes as point particles, expands the metric about Minkowski spacetime
and models retardation in gravitational waves perturbatively. In this paper, we
employ the 2.5PN NZ metric of~\cite{Blanchet:1998vx} for nonspinning black
hole binaries.\footnote{A term of ${\cal{O}}(v^{2N})$ will be said to be of
relative $N$PN order.} 

The FZ is the region sufficiently far from the binary's center of mass that the
metric can be described through a multipolar, PM formalism.
Mathematically, this region can be defined via $r \gg \lambda$ where $r$ is
the distance from the center of mass. In this region,
the metric can be obtained via direct integration of the relaxed Einstein
equations~\cite{Will:1996zj, Pati:2000vt, Pati:2002ux}, or, alternatively, via
the multipolar formalism in~\cite{Blanchet:1995fg, Blanchet:1995fr,
Blanchet:1996wx, Blanchet:2006zz}. Radiative effects are dominant in the FZ and
retardation effects are large. In this paper, we employ a 2.5PN order PM
metric, evaluated explicitly for quasicircular binaries
in~\cite{JohnsonMcDaniel:2009dq}. 

For binaries in the inspiral regime, these different zones overlap in BZs,
shown as cyan shells in Fig.~\ref{fig:zones}. Each IZ overlaps with the NZ,
leading to 2 IZ-NZ BZs, while the NZ overlaps with the FZ in the NZ-FZ BZ.
Inside the IZ-NZ BZ,  the IZ's perturbed Schwarzschild metric and the NZ's PN
metric are both simultaneously valid. Similarly, in the NZ-FZ BZ, the NZ's PN
metric and the FZ's PM metric are simultaneously valid. The existence of such
BZs is what allows one to carry out asymptotic matching and construct a global
metric. 

\subsection{Asymptotic matching}

\subsubsection{Basics and prior work}

Asymptotic matching is the mathematical technique that forces two approximate
solutions to the same set of differential equations to asymptotically approach
each other in their overlapping region of validity. This is achieved through a
given coordinate and parameter transformation that relates the coordinates and
parameters native to each approximation. Such transformations are obtained by
setting the asymptotic expansion of the approximate solutions in the
overlapping regions equal to each other. 

For the problem at hand, asymptotic matching will be used to relate the
perturbed Schwarzschild black hole metric of IZ$A$ to the two-body PN metric of
the NZ. The NZ and FZ metrics are already asymptotically matched by
construction~\cite{Blanchet:2006zz}. Through asymptotic matching, one obtains a
coordinate and parameter transformation to relate IZ$A$ to the NZ, such that
the transformed metric of IZ$A$ asymptotes to the NZ metric in the IZ$A$-NZ
BZ.  

Such a procedure to construct an approximate global metric is also ideal to
compute initial data for binary black hole simulations. Alvi~\cite{Alvi:1999cw,
Alvi:2003pn} was the first to attempt such a construction, but ended up
carrying out {\emph{asymptotic patching}} rather than matching.\footnote{When
patching, one sets the metrics equal to each other at a point, instead of in an
entire BZ region (for more details see~\cite{Yunes:2005nn, Bender}).}  Yunes
{\it et al}.~\cite{Yunes:2005nn, Yunes:2006iw, Yunes:2006mx,
JohnsonMcDaniel:2009dq} succeeded in carrying out matching for nonspinning
binary black holes and this data were recently evolved
in~\cite{Reifenberger:2012yg} (see also~\cite{Chu:thesis} for numerical
evolutions of superposed tidally perturbed BHs). Gallouin {\it et
al}.~\cite{Gallouin:2012kb} recently extended this construction to spinning
black hole binaries.  

One may be tempted to use such a global ``initial data metric'' to obtain one
that is valid on a large number of spatial hypersurfaces. Such a procedure,
however, is doomed to fail from the start. The reason is that \emph{all}
initial data metrics make explicit use of expansions about $t = t_{0}$, where
$t_{0}$ labels the time-parameter of the initial spatial hypersurface. Such
expansions prohibits the use of the initial data metric at $t > t_{0}$,
rendering them useless for dynamical evolutions. One may attempt to
glue several such initial data metrics together to generate a
4-dimensional metric. However, this procedure
will introduce inconsistencies in the 
 coordinate transformations
used in the matching procedure, leading to an invalid metric.

We here take a different route that is guaranteed to produce a global metric
with IZ, NZ and FZ approximate metrics that asymptotically match each other for
all times (up until the binary's orbital separation becomes too small). The
procedure is simply to carry out the matching without assuming 
$(t-t_{0})/r_{12} \ll 1$. Fortunately, part of this matching procedure has
already been carried out by Taylor and Poisson~\cite{Taylor:2008xy}. We will
use the results of this paper, together with the formalism
of~\cite{Yunes:2005nn, Yunes:2006iw, Yunes:2006mx, JohnsonMcDaniel:2009dq},
following closely~\cite{JohnsonMcDaniel:2009dq}, to construct the global
metric.  

We begin by reviewing the work of Taylor and Poisson~\cite{Taylor:2008xy}.
In that paper, they study the geometry of the event horizon of a nonspinning
black hole that is perturbed by a companion in a quasicircular orbit. Their
analysis follows Poisson's approach to black hole perturbation
theory~\cite{Poisson:2003wz,Poisson:2004cw,Poisson:2005pi,Poisson:2009qj,Comeau:2009bz,Poisson:2011nh,Chatziioannou:2012gq},
where the metric close to the black hole (the IZ metric) is expressed in
ingoing Eddington-Finkelstein coordinates. This metric is composed of a
background (the Schwarzschild metric) plus a metric perturbation, expressed as
the product of certain functions of radius, tensor spherical harmonics of the
angular coordinates, and certain functions of time. The latter characterizes
the external universe (the external perturber) and it is expanded in terms of
electric and magnetic tidal tensors. 

Taylor and Poisson~\cite{Taylor:2008xy} asymptotically matched this metric to a
1PN metric in the IZ-NZ BZ. Their matching algorithm is different from the one
pioneered in~\cite{Yunes:2005nn}. In the latter, the matching transformation
consists of both a parameter and a coordinate transformation, which is
essential because the IZ and NZ metrics are not in the same coordinate system
or gauge. Taylor and Poisson~\cite{Taylor:2008xy}, instead, do a series of
coordinate and gauge transformations on the IZ metric so that the end result is
in harmonic coordinates and the perturbation is in harmonic gauge. Once this is
done, the matching transformation just relates the parameters between the two
metrics (i.e., the tidal tensors as functions of the NZ parameters). 

Nowhere in the Taylor and Poisson analysis~\cite{Taylor:2008xy} is the IZ or NZ
metrics expanded in terms of $(t - t_{0})/r_{12} \ll 1$, since they were not interested in
initial data. In that paper, they in fact show by explicit calculation that
asymptotic matching does not require this extra assumption, and thus, the
parameter transformation they find is valid for all times, as long as a BZ
exists. By composing the many coordinate transformations
of~\cite{Taylor:2008xy}, one can also obtain a
time-dependent coordinate transformation that relates the IZ to the NZ metrics. 

One caveat should be mentioned here before proceeding. Taylor and
Poisson~\cite{Taylor:2008xy} only match the metric components that are needed
to study the evolution of the event horizon, namely the shift and the lapse,
but not the spatial metric. Moreover, they work only to 1PN order in the NZ and
to quadrupole order in black hole perturbation theory in the IZ. Therefore,
when applying their scheme to our problem, certain pieces of the 2PN NZ
metric will not properly match to certain pieces of the octupole order IZ metric. 
In order for the global metric to match
consistently to that order, the matching calculation would have to be redone,
which will be considered elsewhere. 

\subsubsection{Time-dependent matching}

We define the order of the matching calculation as follows. Let the IZ
metric be composed of a background, such as the Schwarzschild spacetime, plus
deformations of multipole orders up to $\ell = N+1$, where $N$ is an
arbitrary non-negative integer.   Let the NZ
metric be a PN metric of $N$th PN order. With these definitions, we define the
following:
\begin{itemize}
\item \emph{First-Order Matched Metric}: Constructed from an IZ metric that
consists of the Schwarzschild metric with a quadrupole deformation ($\ell=2$).
This perturbation depends only on the electric quadrupole ${\cal{E}}_{ab}$ and
the magnetic quadrupole ${\cal{B}}_{ab}$. The NZ metric consists of a 1PN
metric. 
\item \emph{Second-Order Matched Metric}: Constructed from an IZ metric that
consists of the Schwarzschild metric with a quadrupole and an octupole
deformation ($\ell=2$ and $\ell =3$). This perturbation depends on the electric
quadrupole ${\cal{E}}_{ab}$ and the magnetic quadrupole ${\cal{B}}_{ab}$, as
well as on their time derivatives $\dot{{\cal{E}}}_{ab}$ and
$\dot{{\cal{B}}}_{ab}$, and the electric octupole ${\cal{E}}_{abc}$  and the
magnetic octupole ${\cal{B}}_{abc}$. The NZ metric consists of a 2PN metric.  
\end{itemize}
Of course, nothing prevents us from using the most accurate PN metric in the NZ
that is available, possibly resummed in some way. However, this does not imply
the IZ metric will asymptotically match the NZ metric to higher order.  

The global metrics will be constructed as follows. We take the results
of~\cite{JohnsonMcDaniel:2009dq}, which are formally valid only about a given
spatial hypersurface, and perform a ``temporal resummation''\footnote{Resummation is a mathematical technique
whereby a (possibly divergent) series expansion is replaced by a single function,
whose Taylor expansion is identical to the original series. Many types of resummations exist
in the literature, such as Pad\'e resummation~\cite{Bender}. Temporal resummation amounts
to a particular trigonometric resummation on the time-variable only.}: we replace terms 
that depend on powers of $\omega t$ in the
coordinate and parameter transformation of~\cite{JohnsonMcDaniel:2009dq}, with $\omega$ the orbital frequency, by
functions of the PN orbital trajectories and velocities. 
For example, a term of the form $\omega t$ can be temporally resummed into $\sin {\omega t}$, which is
proportional to the $y$ component of the trajectory of body 1. Similarly, a term of the form
$-\omega^{2} t$ can be temporally resummed into $-\omega \sin{\omega t}$, which
is proportional to the $x$ component of the velocity of body 1.  

Note that the operation of
temporal resummation is not unique. For example, a term of the form $\omega^{2}
t^{2}$ can be temporally resummed either by $\sin^{2}{\omega t}$ or by $2 (1 -
\cos{\omega t})$. So how does one choose the proper resummation?  Here
we will be guided by the work of Taylor and Poisson~\cite{Taylor:2008xy}.
We cannot use their transformations directly because they work in a different
gauge and coordinate system. However, we can use the tensorial structure of
their results as guiding principles to properly carry out the temporal
resummation. As we will show next, using this insight, we can temporally resum
the work of~\cite{JohnsonMcDaniel:2009dq} to obtain an IZ metric that formally
asymptotically matches the NZ metric to first order in all components and for
all times.  As explained in the caveat of the last subsection, we cannot use
the work of Taylor and Poison to gain insight into temporal resummation at
second-order in matching, since they work to first-order. Second-order,
temporally resummed, asymptotic matching will have to be derived from first
principles, as we will study in a future paper. 

\subsubsection{First-order matching}

Based on~\cite{Taylor:2008xy}, we here resum the time dependence of the
coordinate transformation found in~\cite{JohnsonMcDaniel:2009dq}. First, we
prepare various functions, as motivated by~\cite{Taylor:2008xy}.  Let us define
$\tilde{x}^i = \{\tilde{x},\,\tilde{y},\,\tilde{z}\}$ as the coordinates
centered on BH1,
\bea
\tilde{x} &=& x-{\frac {{m_2}\,{r_{12}}\,}{M}} \cos{\phi} \,,
\\
\tilde{y} &=& y-{\frac {{m_2}\,{r_{12}}\,}{M}} \sin{\phi} \,,
\\
\tilde{z} &=& z \,,
\eea
where  $\tilde{r} = \sqrt
{{{\tilde{x}}}^{2}+{{\tilde{y}}}^{2}+{{\tilde{z}}}^{2}}$ is the radial 
coordinate from BH1. 
The orbital phase evolution $\phi = \phi(t)$ and the evolution of the orbital separation $r_{12} = r_{12}(t)$ can be calculated in
the PN formalism and must be evaluated in the same way as in the NZ PN metric.
Keep in mind, however, that these quantities are both time dependent, and thus
their temporal evolution must be included when computing the Jacobian of the
coordinate transformation.  

Let us also define the coordinates centered on BH1 that are 
corotating with the binary:
\bea
\tilde{x}_c &=& {\tilde{x}} \; \cos{\phi}\,+ {\tilde{y}} \; \sin{\phi} \,,
\cr
\tilde{y}_c &=& {\tilde{y}} \; \cos{\phi}\, - {\tilde{x}} \; \sin{\phi} \,.
\eea
These quantities arise from the inner products $\tilde{x}^i \hat{x}_{(p)i}$ and
$\tilde{x}^i \hat{v}_{(p)i}$ respectively, where $\hat{x}_{(p)i}$ and
$\hat{v}_{(p)i}$ denote the unit vectors of the PN particles' locations from the
center of mass and their velocities. Note also that $\tilde{x}_c \sim {\tilde{x}}
+ \phi(t) \,{\tilde{y}}$ and $\tilde{y}_c \sim {\tilde{y}} - \phi(t)
\,{\tilde{x}}$ for small $\phi(t) \ll 1$. If we neglect radiation-reaction,
$\phi(t) = \omega t = v_{12} t/r_{12}$, so  small $\phi(t)$ means an expansion
in $t/r_{12} \ll 1$, precisely the expansion used
in~\cite{JohnsonMcDaniel:2009dq}.

Let us now follow Taylor and Poisson~\cite{Taylor:2008xy} and introduce
quantities similar to their Eqs.~(7.21) and (7.22):
\bea
\dot{A} &=& \frac{1}{2}\,{\frac {{m_2}\, \left( {m_2}+2\,M \right) }{M \; {r_{12}}}} \,,
\cr
R^i &=& \{0,\,0,\,R^z\} \,,
\cr
R^z &=&
-\frac{1}{2}\,{\frac {{m_2}\, \left( 4\,M-{m_2} \right) \sin{\phi}\, \cos{\phi}}
{{M}\,{{r_{12}}} (1+(m_2-m_2^2/M-3M)/(2 r_{12}))}}
\,.
\eea
The vector $R^i$ shown above is obtained by simple integration, $R^i = \int dt
\dot{R}^i = \dot{R}^i t$, where $\dot{R}^{i}$ is given in~\cite{Taylor:2008xy},
and the last equality holds because $\dot{R}^i$ is constant in time, when one
neglects radiation reaction. Of course, direct integration of $\dot{R}^{i}$
in~\cite{Taylor:2008xy} would lead to a term proportional to $\omega t$ in the
spatial sector of the coordinate transformation, which is unacceptable and has
here been temporally resummed. We empirically find that resumming $\omega t$
through $\sin{\phi} \cos{\phi}$ leads to proper matching at first order. Of
course, the resummed form of $R^{z}$ shown above reduces exactly to the
Taylor-Poisson expression for $\omega t \ll 1$. On the other hand, we
temporally resum $A = \int dt \dot{A} = \int dr_{12} \dot{A} (dr_{12}/dt)^{-1}$,
where the rate of change of the orbital separation $dr_{12}/dt$ is calculated
from the balance law (for example, see~\cite{Blanchet:2006zz}), because this
term enters the time component of the coordinate transformation only, which
evolves in the radiation reaction timescale, as we will see below. We could have
resummed $R^{i}$ similarly to $A$, but we find that this is not necessary to the
order we work here.

Then, taking into account the form of the coordinate transformation in
Ref.~\cite{Taylor:2008xy}, the following transformation is obtained
as an extension of that in Ref.~\cite{JohnsonMcDaniel:2009dq}:
\begin{widetext}
\begin{align}
T &= 
t-{\frac {{m_2}\,{\tilde{y}_c}}{\sqrt {{r_{12}}}\sqrt {M}}}
+\frac{5}{384} \frac{(m_2+2 M)(r_{12}^3-r_{12}(t_0)^3)}{M^2 m_1}
+\biggl[ -\frac{1}{2}\,{\frac {{m_2}\,{\tilde{y}}\,\sqrt {M}{\tilde{x}}\,{{\tilde{r}}}^{2}}{{{r_{12}}}^{9/2}}}
+\frac{1}{2}\,{\frac {{m_2}\,{\tilde{y}_c}\,\sqrt {M}{{\tilde{r}}}^{2}}{{{r_{12}}}^{7/2}}}
+\frac{5}{2}\,{\frac {{m_2}\,{\tilde{y}}\,{{\tilde{x}}}^{3}\sqrt {M}}{{{r_{12}}}^{9/2}}}
\cr & \quad
-2\,{\frac {{{\tilde{x}}}^{2}\sqrt {M}{\tilde{y}}\,{m_2}}{{{r_{12}}}^{7/2}}}
-{\frac {{m_2}\,{\tilde{y}_c}\,{\tilde{x}_c}\, \left( {m_2}-2\,M \right) }{\sqrt {M}{{r_{12}}}^{5/2}}}
+\frac{1}{2}\,{\frac {{m_2}\,{\tilde{y}_c}\, \left( -5\,M+{m_2} \right) }{\sqrt {M}{{r_{12}}}^{3/2}}} \biggr] \,,
\cr
X &=
{\tilde{x}}
+\left[ -\frac{1}{2}\,{\frac {{{m_2}}^{2}\sin{\phi}\,{\tilde{y}_c}}{M{r_{12}}}}
+{\tilde{x}}\, \left( {\dot{A}}-\frac{1}{2}\,{\frac {{{m_2}}^{2}}{M{r_{12}}}} \right) 
+\frac{1}{2}\,{\frac {{m_2}\, \left( -2\,{\tilde{x}}\,{\tilde{x}_c}+\cos{\phi}\,{{\tilde{r}}}^{2} \right) }{{{r_{12}}}^{2}}} \right] 
-{\tilde{y}}\,{R^z} 
\,,
\cr
Y &= 
{\tilde{y}}
+\left[ \frac{1}{2}\,{\frac {{{m_2}}^{2}\cos{\phi}\,{\tilde{y}_c}}{M{r_{12}}}}
+{\tilde{y}}\, \left( {\dot{A}}-\frac{1}{2}\,{\frac {{{m_2}}^{2}}{M{r_{12}}}} \right) 
+\frac{1}{2}\,{\frac {{m_2}\, \left( -2\,{\tilde{y}}\,{\tilde{x}_c}+\sin{\phi}\,{{\tilde{r}}}^{2} \right) }{{{r_{12}}}^{2}}} \right]
+{\tilde{x}}\,{R^z}
 \,,
\cr
Z &=
z
+ \left[ z \left( {\dot{A}}-\frac{1}{2}\,{\frac {{{m_2}}^{2}}{M{r_{12}}}} \right) 
-{\frac {{m_2}\,z{\tilde{x}_c}}{{{r_{12}}}^{2}}} \right] 
 \,,
 \label{coord-transf}
\end{align}
\end{widetext}
The IZ1 metric components are expressed in terms of the coordinates $\{T,X,Y,Z\}$;  
the transformation from $\{t,x,y,z\}$ to $\{T,X,Y,Z\}$, therefore, provides a means 
by which we can relate points in the NZ to those in the IZ1 (with similar transformations
for IZ2 after the $1 \leftrightarrow 2$ exchange symmetry transformation). 
As mentioned earlier, notice that the temporal piece of the transformation
contains the orbital separation, $r_{12}(t_0)$, at a point in time, $t_0$, to
describe the orbital evolution used to calculate the radiation reaction timescale; 
we set $t_0$ to be the initial time of a simulation. 

Note that we have only discussed the resummation of the coordinate transformation
and have not mentioned resumming the parameter transformation. The
pieces of this transformation that are needed to be resummed are the relations
of the multipole tidal fields as functions of the PN parameters. Such temporal
resummation was already carried out in the Appendices
of~\cite{JohnsonMcDaniel:2009dq}, and will thus not be presented again here.

With this resummed coordinate and parameter transformation, one can carry out
several tests to check whether the temporal resummation was successful. First,
we checked that the above transformation agrees \emph{exactly} with that
of~\cite{JohnsonMcDaniel:2009dq} in the $\omega t \ll 1$ limit, 
i.e., when $t/r_{12} \ll 1$. This automatically implies that the IZ and NZ
metrics matches to first order in the BZ for $t \ll r_{12}$. Second, we
evaluated the transformation at a point on the horizon and plotted it as a
function of time. We found that the transformation above takes this point to a
trajectory identical to that of the NZ PN point particles, as expected. Third,
we asymptotically expanded the transformed first-order IZ metric in the BZ and
the NZ 1PN metric in the BZ (both without expanding in $t/r_{12} \ll 1$). We
then compared every single metric component and found that they were identical in the
BZ. This then proves that the temporally resummed transformation leads to
first-order asymptotic matching for all times (as long as a BZ exists).

Before proceeding, we should mention that the transformation above is
technically different from that in~\cite{Taylor:2008xy}. In particular, we did
not use the contribution to the transformation presented in their Eq.~(6.16),
and we changed the sign of the second term in the right-hand side of their
Eq.~(5.5). Their transformation and ours need not agree because Taylor and
Poisson start with an IZ metric in a different gauge and coordinate system than
the one used here. The functional form of the transformation, however, is
indeed the same, which was crucial to select the proper temporal resummation.
Ultimately, what really matters for our purposes is that the above
transformation has been found to satisfy the asymptotic matching equations to
first order. 

\subsubsection{Second-order matching}

At second order in the matching, the problem becomes \emph{dramatically} more
complicated. Obviously, part of the complication is that the metrics in the IZ
and NZ are themselves longer and more complicated (2PN versus 1PN, octupole
deformation versus quadrupole). Symbolic manipulation software, such as Maple,
can barely handle the necessary calculations presented here with our
computational resources. But besides this, the main problem with temporally
resumming the second-order matching transformation is that we have no guidance
as to how to properly do it, because Taylor and Poisson only worked to
first order. Without such guidance, there is an infinite number of ways in
which the resummation can be carried out and not all will actually lead to a
properly, second-order matched global metric.

In spite of these difficulties, we will proceed and attempt to temporally resum
the transformation of~\cite{JohnsonMcDaniel:2009dq} at second order in
matching. First, let us take the temporally resummed transformation in
Eq.~\eqref{coord-transf} and expand it in $t/r_{12} \ll 1$. Comparing this to
the second-order transformation in~\cite{JohnsonMcDaniel:2009dq}, one finds
that the transformations do not agree. More in detail, the latter contains
terms that do not arise upon expanding Eq.~\eqref{coord-transf} in $t/r_{12}
\ll 1$ {\emph{and}}, the expanded version of Eq.~\eqref{coord-transf} generated
terms that are not contained in the second-order transformation
of~\cite{JohnsonMcDaniel:2009dq}. Clearly, additional terms must be added to
Eq.~\eqref{coord-transf} to properly match at second order.  

Let us then try to temporally resum the difference between the second-order
transformation of~\cite{JohnsonMcDaniel:2009dq} and the $t/r_{12}$-expanded
version of Eq.~\eqref{coord-transf}. Using some guidance from the first-order
matching calculation (but of course, as explained above, this guidance is
limited), one can temporally resum the difference. The result is truly
formidable, and thus, we will show it in Appendix~\ref{app:equations}. 
Adding this resummed difference to Eq.~\eqref{coord-transf} then provides {\emph{a}} full,
temporally resummed coordinate transformation to second-order in the matching.
But of course, we have no guarantee that this particular choice of temporal resummation is
the correct one, i.e., the one that leads to full asymptotic matching at second order for all times. 

As in the first-order matching case, we can analytically investigate this last issue to see
how well the second-order metric performs. First, we checked that the
temporally resumed second-order transformation agrees {\emph{exactly}} with
that of~\cite{JohnsonMcDaniel:2009dq} in the $\omega t \ll 1$ limit,
i.e., when $t/r_{12} \ll 1$. As before, this automatically implies that the
the IZ and NZ metrics match to second-order in the BZ at $t \ll r_{12}$.
Second, we evaluated the transformation at a point on the horizon and plotted
it as a function of time. Again, we found that the second-order transformation
takes this point to a trajectory identical to that of the NZ PN point
particles.  Third, and perhaps most importantly, we asymptotically expanded the transformed IZ metric
in the BZ and the NZ metric in the BZ (both without expanding in $t/r_{12} \ll
1$). This time, however, we did not find that all metric components matched in
the BZ. That is, the temporally resummed second-order metric is not properly 
asymptotically matched for all times.   

\begin{figure*}[htb]
\begin{center}
\includegraphics[width=0.45\textwidth,clip=true]{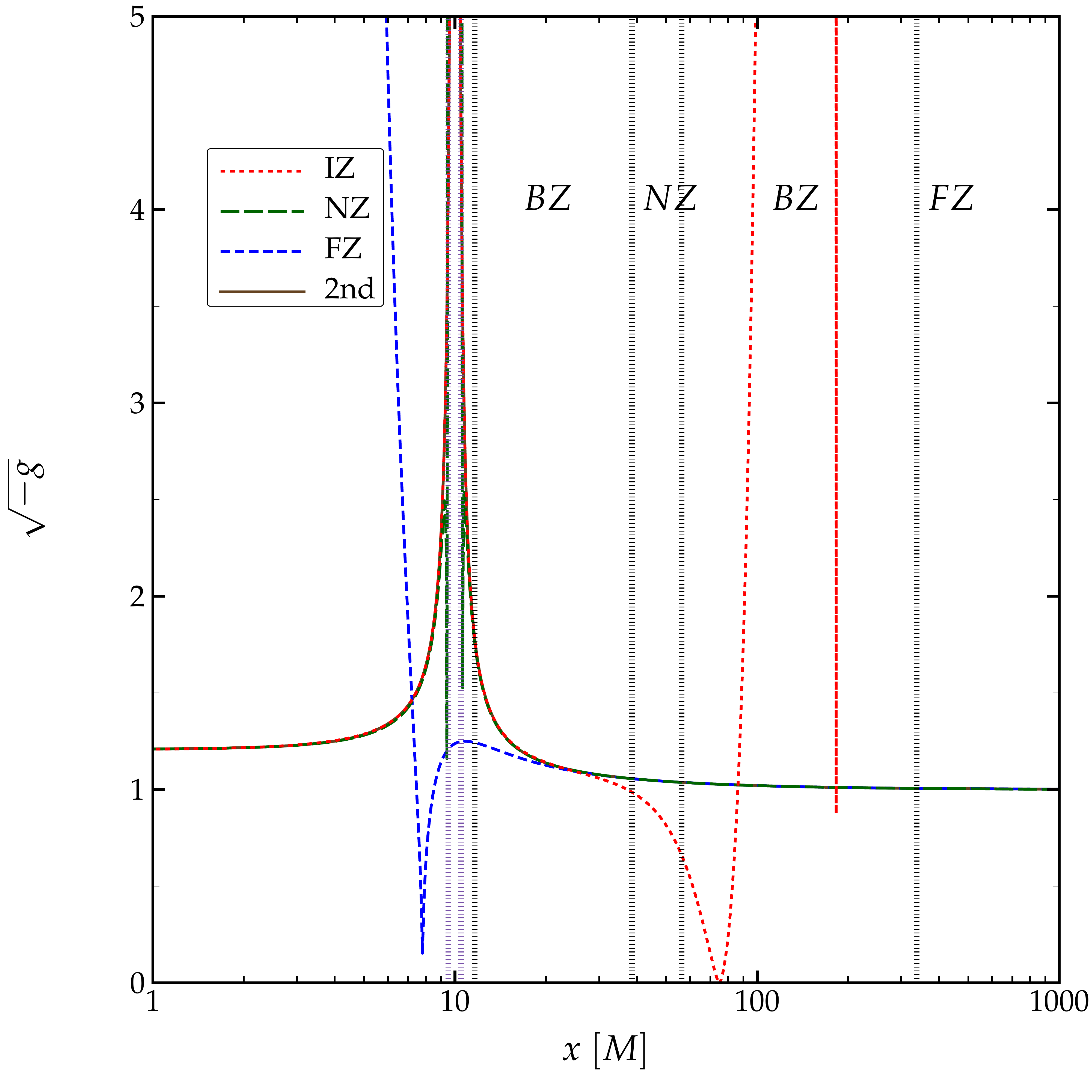}
\includegraphics[width=0.45\textwidth,clip=true]{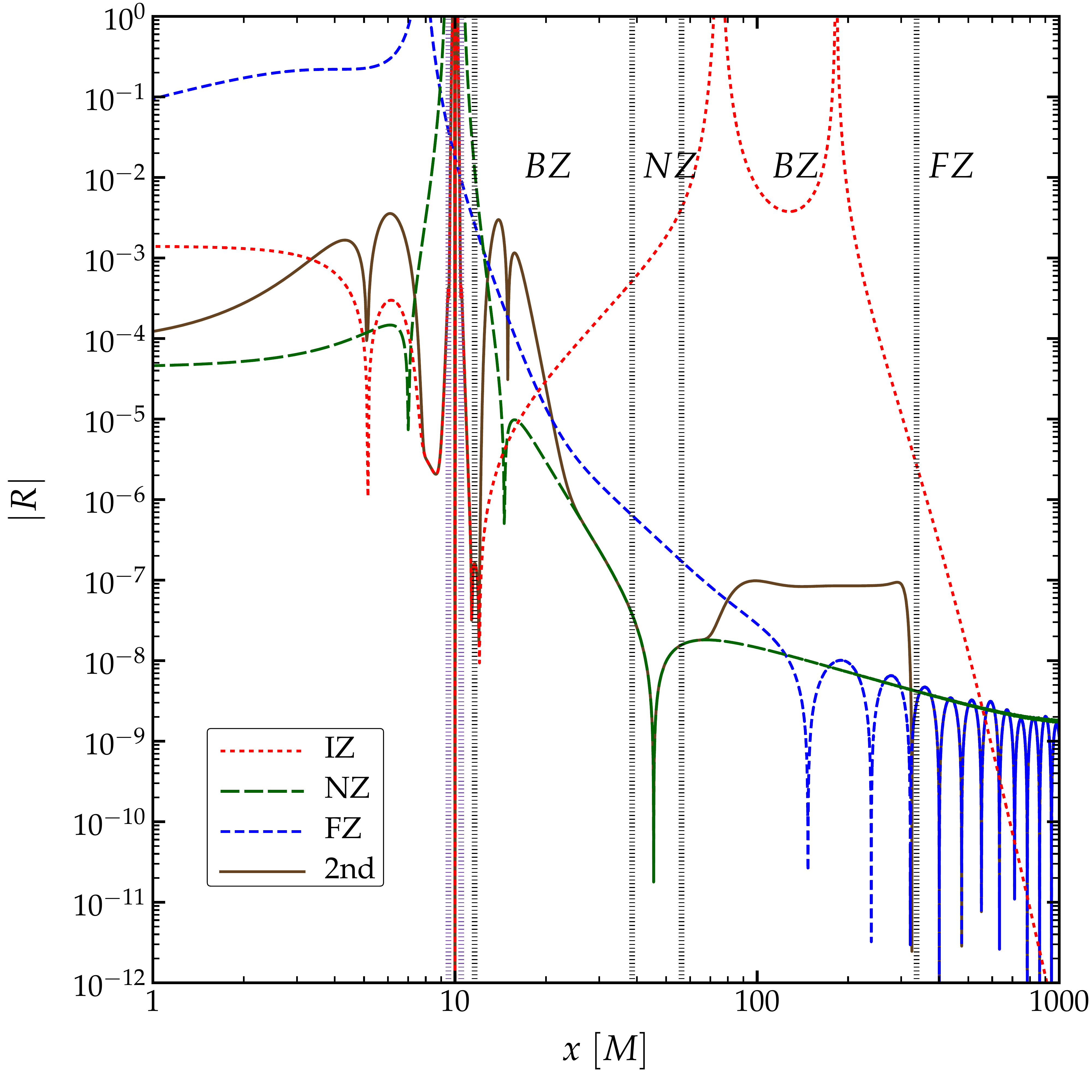}
\end{center}
\caption{(Color online) Left: Comparison of the volume element, $\sqrt{-g}$, 
for the several different metric pieces composing the global, second-order
matched metric, viz., the IZ, the NZ and the FZ metrics. Right: Absolute value
of Ricci scalar computed with these metric pieces. 
The vertical dotted lines mark the boundaries of each zone.}
\label{fig:metric_pieces}
\end{figure*}

For example, the off-diagonal spatial parts of the
metric (e.g.~$g_{12}$, $g_{13}$ and $g_{23}$) do not match at second
order. At first order in the
matching, one needs not worry about these components, as they simply do not
enter the 1PN metric or the transformed IZ metric. Of course, at second-order
in the matching, the 2PN NZ metric does have nonzero off-diagonal,
spatial-spatial components, and these must be properly matched by the
transformed IZ metric. In order to achieve this, we modified the resummed,
second-order transformation so that many of the terms in the $(x,\,y)$
component would match. This required not only adding terms at ${\cal{O}}(v^4)$
to the transformation, but also some at
${\cal{O}}(v^5)$~\cite{JohnsonMcDaniel:2009dq}. Given this analysis, it would
seem that to complete the matching at second-order for all times,
${\cal{O}}(v^6)$ pieces may be needed in the $t/r_{12} \ll 1$ expanded
transformation, which are currently unknown. Needless to say, in spite of this
improvement, the $(y,z)$ or $(x,z)$ components continue not to match at second
order.    

One may here wonder whether this second-order, temporally resummed matching
transformation is better than the first-order one, since after all it does not properly
lead to asymptotic matching in all components for all times. Evaluating the metric
components, however, we empirically find that the second-order transformed IZ metric is
actually much closer to the NZ metric in the BZ for all time. The improvement
is roughly a factor of 5 relative to the first-order matching transformation.
Of course, we suspect that if second-order, time-dependent matching were
carried out properly and from first principles, the improvement between first
and second-order matching would be even better (perhaps a factor of 7 or 10),
but such an analysis will have to wait to future work. 

\subsection{Global metric}

Once the IZ metrics have been transformed with the first or second-order,
temporally resummed transformations, one must still glue each IZ metric to the NZ metric
with proper transition functions. The global metric is then simply the weighted
average
\begin{eqnarray}
 g_{\mu\nu} &=&
 (1 - f_{\rm far})
 \Bigl\{f_{\rm near} \bigl[f_{{\rm inner},1} \,g_{\mu\nu}^{\rm (NZ)} 
 +(1 - f_{{\rm inner},1} ) \,g_{\mu\nu}^{\rm (IZ1)}\bigr]
 \nonumber \\ &&
 + (1 - f_{\rm near} )\bigl[f_{{\rm inner},2} \,g_{\mu\nu}^{\rm (NZ)} 
 +(1 - f_{{\rm inner},2} ) \,g_{\mu\nu}^{\rm (IZ2)}\bigr]\Bigr\}
 \nonumber \\ &&
 + f_{\rm far} \,g_{\mu\nu}^{\rm (FZ)} \,,
 \label{eq:wholemetric}
\end{eqnarray}
where the transition functions, $f_{\rm far}$, $f_{\rm near}$, $f_{{\rm
inner},1}$ and $f_{{\rm inner},2}$ are summarized in Appendix~\ref{app:trans}.
The Maple scripts and C codes for the global, IZ, NZ and FZ metrics are all
freely available online as ``Supplemental Material''
in~\cite{JohnsonMcDaniel:2009dq}. We have modified these scripts to include the
temporally resummed matching transformation and to optimize them for speed. 

Figure~\ref{fig:metric_pieces} shows how the several metric pieces
contribute to the global second-order matched metric. Observe on the left panel
how smoothly the IZ and NZ metric pieces approach each other 
in the BZ to make up the global metric. Furthermore, note
on the right panel how each metric piece leads to different violations of the
Ricci scalar. In particular violations outside the validity region of the
several metric pieces become much larger than the ones resulting from the
global metric.

\section{NUMERICAL ANALYSIS}
\label{sec:ricci}

Using the metric in Eq.~(\ref{eq:wholemetric}), we calculate the Einstein
tensor $G_{\mu\nu}$ (or Ricci tensor $R_{\mu\nu}$, Ricci scalar $R$,
Hamiltonian constraint ${\cal H}$, and momentum constraint ${\cal M}_{i}$) to determine
the accuracy of our approximations.  Our
computation does not depend on the Arnowitt-Deser-Misner (ADM, or $3+1$)
decomposition~\cite{Arnowitt:1962hi}, so we emphasize four dimensional
quantities in our analysis below.  We will here show that the Einstein tensor
vanishes to the appropriate order not only at an initial moment of time, but on
a sequence of time slices or arbitrary spatial extent, as expected.

\subsection{Basic equations}

An exact binary black hole vacuum spacetime must satisfy the ten vacuum Einstein
equations $R_{\mu \nu}=0$.  Hence deviations of $R=g^{\mu \nu} R_{\mu \nu}$
from zero are a measure of the error introduced in our analytical construction.
We here use the conventions in~\cite{MTW} to evaluate the Ricci scalar.
Another approach to estimate the error on a given time slice 
is to calculate the violation of the Hamiltonian and momentum constraints. 
We follow here the standard conventions,
for example in~\cite{1987PThPS..90....1N}, to compute these quantities.

Traditionally, the numerical relativity community evaluates the quality and
correctness of the numerical solutions to the Einstein equations 
by monitoring the violation of the Hamiltonian and momentum constraints, 
as a function of space,  time, and resolution.
In analytical relativity, however, one usually computes the Einstein tensor to
determine the correctness of approximate solutions. We have calculated
all of these quantities and found that they present similar behavior. 
In this section, we present the Ricci scalar as a measure of the accuracy of the global
metric.

But how should one interpret this measure?  In vacuum GR,
the Ricci scalar must vanish identically. The global metric constructed here,
however, is approximate, and thus this scalar does not vanish identically.
Physically, this error can be interpreted as a (possibly
energy-condition-violating) nonvacuum component of spacetime, such as a
noncollisional fluid or dust, since the Ricci scalar must be equal to the
trace of the stress-energy tensor.  

One would like to assess if this error is small enough to be negligible
relative to other errors contained in the modeling of the system. This, of
course, depends on the problem one is considering.  In this paper, the problem
one would eventually like to solve is the evolution of a black hole binary with
a circumbinary accretion disk. It is thus natural to compare this error to the
total mass present in the domain, which is dominated by the black hole masses
in the IZs and the NZ.  We plan on exploring further this interpretation in 
future work, computing the total amount of ``fake'' mass that would
be felt by a magnetized disk as a function of the distance to the binary's
center of mass. 

Before presenting a numerical analysis of the above calculations, we give rough
estimates of the location of the horizon and the innermost stable circular
orbit (ISCO).  This will help us understand the numerical result better.  For a
single Schwarzschild black hole with mass $m$, the event horizon is located at
$r_{\rm Hor, Sch}=2m$ and $r_{\rm Hor, Harm}=m$ in Schwarzschild and PN
harmonic coordinates respectively, while the ISCO of a test particle in such a 
spacetime is at $r_{\rm ISCO, Sch}=6m$
and $r_{\rm ISCO, Harm}=5m$ respectively.  Therefore, very roughly, the event
horizon of the global metric for a binary system of comparable masses is
located at $r_{A, {\rm Hor}}=0.5M$ for $A \in (1,2)$ and the ISCO is at $r_{A,
{\rm ISCO}}=2.5M$ in PN harmonic coordinates, where $M=m_1+m_2$ is the total
mass.

\subsection{Numerical method and convergence tests}

Our global approximate metric and associated analysis routines were implemented
and tested both in a standalone code and in the \harm code.  The former allowed
us to perform point-wise tests of the routines anywhere on the $4$-dimensional
manifold. The latter provided a platform to efficiently study the routines as a
time series of $3$-dimensional slices, appropriately chosen to cover the
regions around each black hole.  \harm~\cite{Noble09,Noble10,Noble:2011wa} is a
GRMHD code originally developed to study accretion disks around single
black holes. It has recently been redesigned to handle any kind of analytic
metric, including black hole binary metrics such as ours.  For a review of its
capabilities, please refer to~\cite{Noble:2012xz}.

\begin{figure*}[htb]
\begin{center}
\includegraphics[width=0.45\textwidth,clip=true]{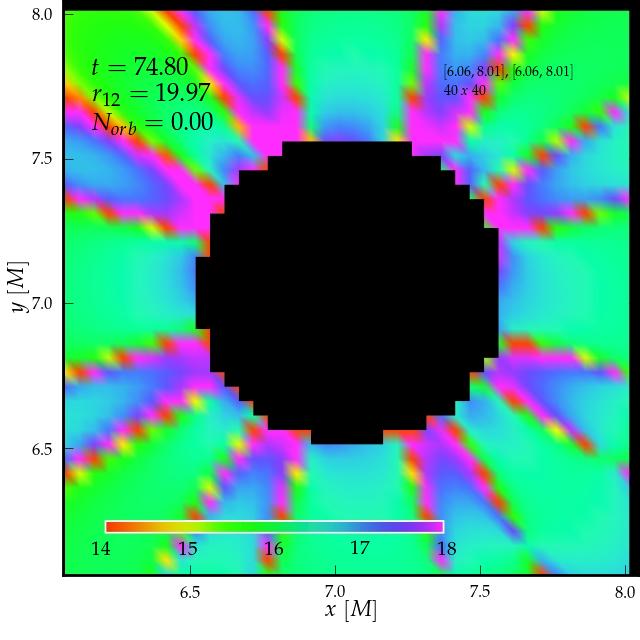}
\includegraphics[width=0.45\textwidth,clip=true]{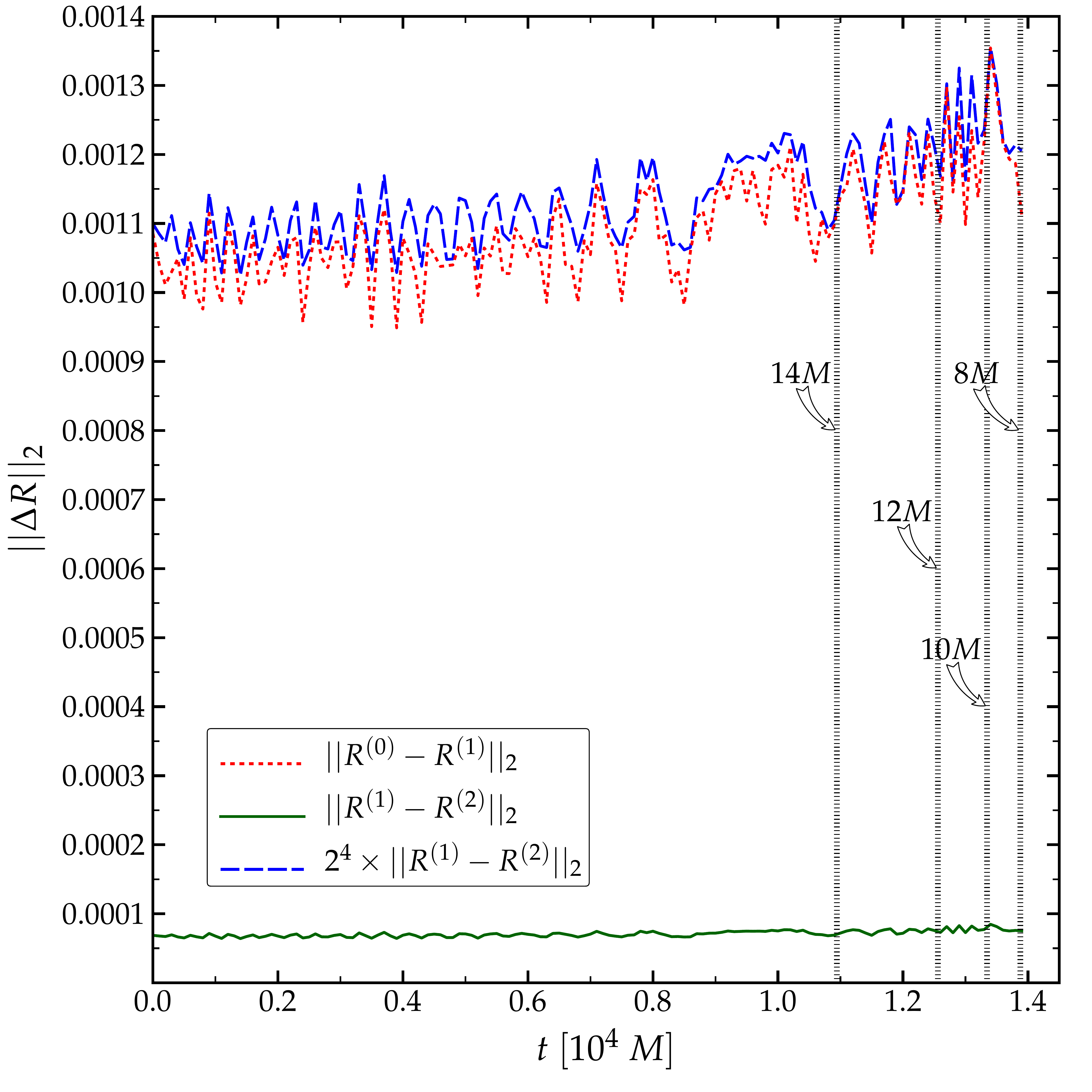}
\end{center}
\caption{(Color online) 
Left: Convergence factor $Q^{h}$ computed using three different
grid resolutions: $h=0.05M$, $h=0.025M$ and $h=0.0125M$, respectively,
corresponding to $10$, $20$ and $40$ points across the horizon radius.  Observe
that this factor is almost everywhere close to $16$, which corresponds to
fourth-order convergence. The magenta regions in this panel show areas that
converge faster that fourth-order, while red regions show areas that converge
slower, possibly due to zero-crossings of the solution error function.  Right: 
L2 norm of the Ricci scalar difference for different levels of
refinement as a function of time. The L2 norm is calculated over a sector of
the simulation domain that ranges from $[-12,-12,-0.75]$ to $[12,12,0.75]$ with
uniform resolutions following a $2:1$ Cartesian grid spacing ratio, where the
coarsest grid has a mesh spacing of $0.05M$.
}\label{fig:2D_GLO2R_zoom_convergence}
\end{figure*}

While the metric used in this paper is purely analytic, we chose to calculate its 
first and second derivatives numerically when computing the Einstein
equations.  We used fourth order finite difference approximations to the
continuum derivative operators in a Cartesian coordinate system. Several
different resolutions were employed to make sure the quantities presented are
in the convergence regime to the continuum solution. We have avoided
implementing an analytic version of the metric derivatives mainly due to the
analytical complexity of the metric components themselves. An implementation 
of the analytic representation of second order mixed derivatives in time and space, for
example, would not be efficient or advantageous from a computational point of view.

In our battery of tests, the first and most fundamental one is related to the
convergence of the numerical solution to the analytic solution.  Recall that we
employed fourth-order, finite-difference approximations to the derivative
operators. One way to assess the convergence rate is to look at the local
convergence factor $Q^h(t)$ for a discrete solution function $u^{h}$ defined
as
\begin{equation} 
Q^h(t) = \frac{u^{4h}-u^{2h}}{u^{2h}-u^{h}} = 2^p + O(h),
\end{equation} 
where $p$ is the order of the finite-difference approximation employed ($p=4$
here). 

Let us now prove that our numerical implementation of Ricci scalar does 
indeed converge to its analytic nonzero value to fourth order, locally and 
globally in time. 
The left panel of Fig.~\ref{fig:2D_GLO2R_zoom_convergence} shows the
convergence factor in a linear color scale ranging from $14$ to $18$.  This
figure shows that most of $Q^h$ stays in that range, suggesting fourth-order
convergence.  The star-ray pattern (regions where the convergence factor
saturates the color scale) corresponds to regions where the discrete-solution,
error function crosses zero and, therefore, cannot be appropriately represented
with finite (double) machine precision. The right panel of the same figure
shows the convergence of the Ricci scalar as a function of time. In particular,
this figure shows the $L2$ norm of the difference of the Ricci scalar evaluated with
different levels of resolution.  The factor between the two curves,
corresponding to the difference between medium and high levels of resolution
and between low and medium ones, is close to $2^4$ as one can verify visually.  
Both panels reassure us of the correctness of our
finite difference implementation. 

\subsection{Accuracy of global metric}

\begin{figure*}[htb]
\begin{center}
\includegraphics[width=0.45\textwidth,clip=true]{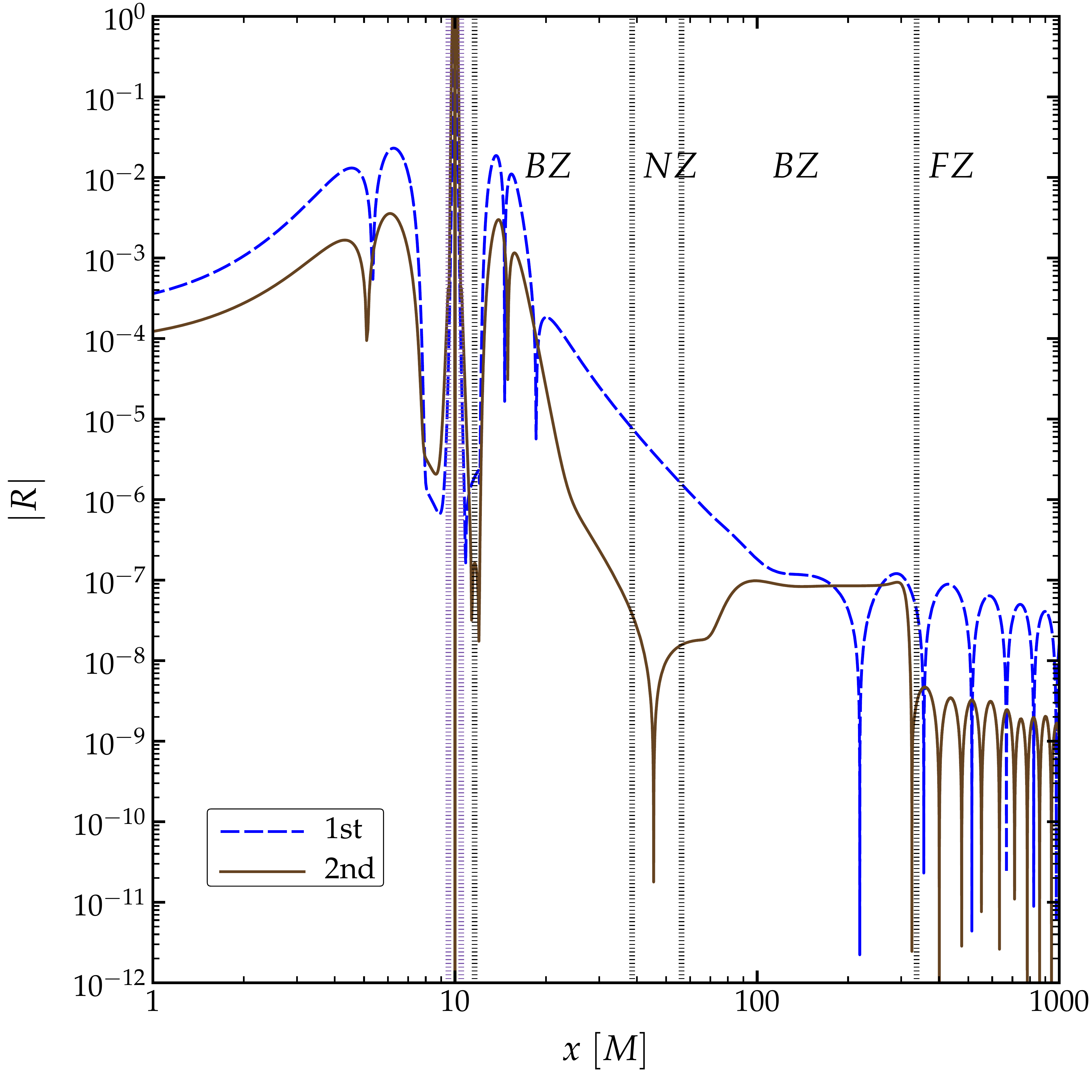}
\includegraphics[width=0.45\textwidth,clip=true]{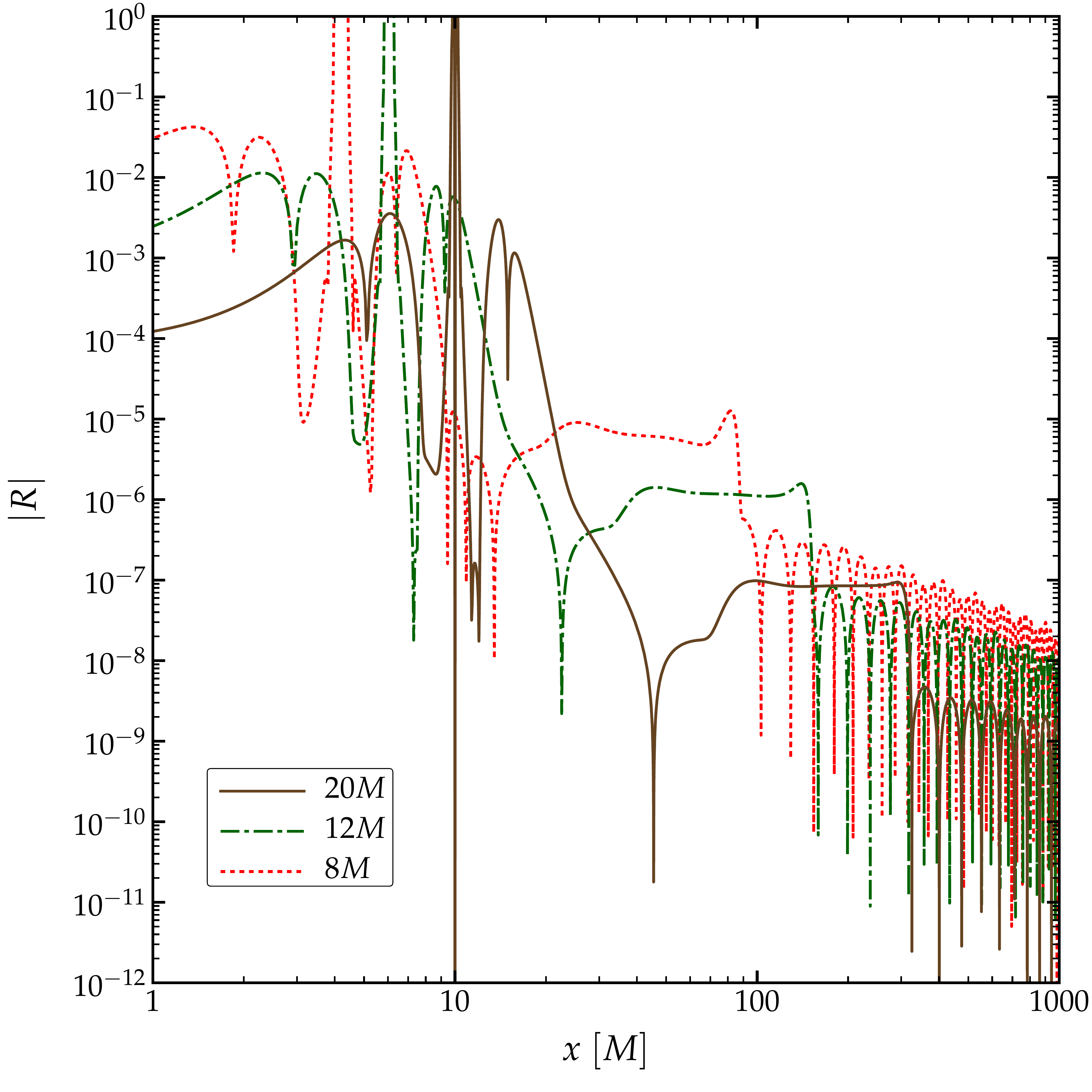}
\end{center}
\caption{(Color online) 
Absolute value of the Ricci scalar along the $x$ axis (left) using the first and second-order metrics at a separation of $20M$ and (right) using the second-order metric at different separations. The mesh spacing is $dx=0.0125M$, roughly $40$ grid points along the horizon radius (unless indicated otherwise, the same
resolution is used in all other figures). The thick purple vertical lines show
the location of the event horizon, while the black ones show the boundary of 
the different zones. Observe how smoothly the violations to the Einstein equations increase as the orbital separation shrinks. }
\label{fig:xcut_R_4th}
\end{figure*}

We turn our attention next to how the Ricci scalar behaves as a function of the
perturbative approximation order of the global metric and as a function of
time. When we introduced our time resummation procedure to adapt the initial
data metric to a global metric, we could have introduced artifacts or
overlooked matching assumptions that could have led to undesirable errors. Figure~\ref{fig:xcut_R_4th} addresses these concerns by showing the Ricci scalar along the $x$ axis. The left panel shows this scalar computed with the first and the second-order metrics at a separation of $20M$. The right panel shows the scalar computed with the second-order metric at several instances of time in the evolution, starting from a separation of $20M$ and ending at $8M$. The different
spacetime zones are marked with vertical lines. 
Observe that the magnitude of the Ricci scalar
decreases as the order of the approximation increases. Furthermore, the violations  increase slowly and smoothly as the orbital separation shrinks. These figures show that no artifacts or pathologies have been introduced in the global metric.

\begin{figure*}[htb]
\begin{center}
\includegraphics[width=0.45\textwidth,clip=true]{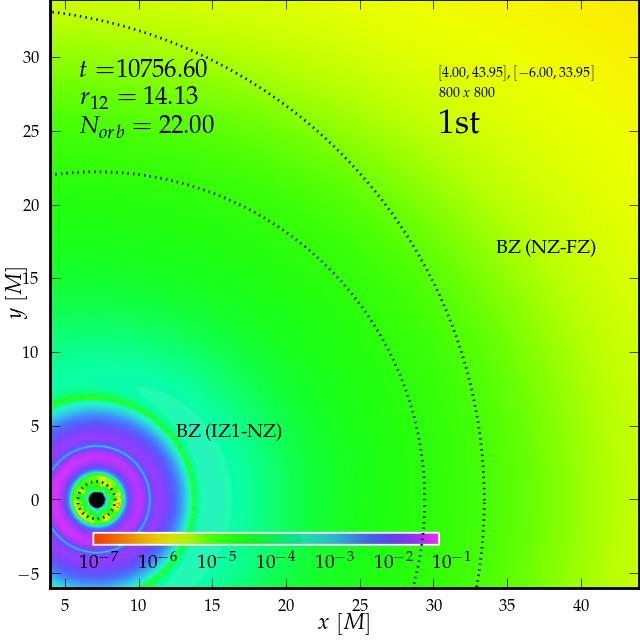}
\includegraphics[width=0.45\textwidth,clip=true]{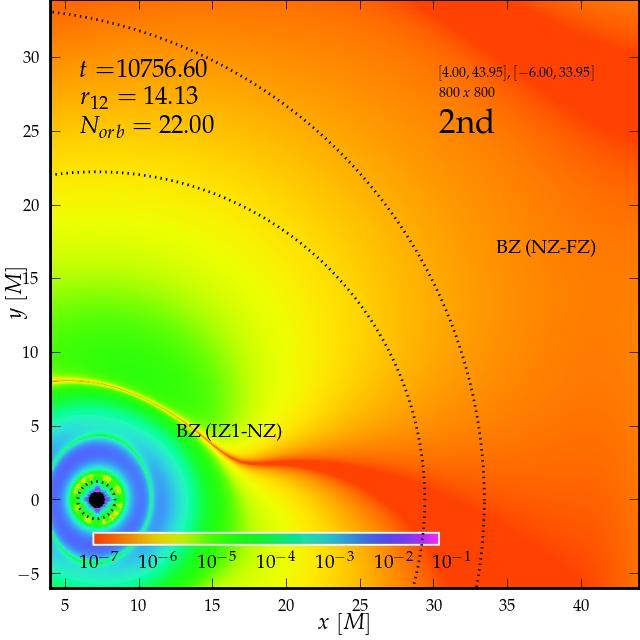}
\end{center}
\caption{(Color online) 
Comparison of the absolute value of the Ricci scalar calculated with the first order
(left) and the second order (right) metrics at $t=10756.60 M$, corresponding to an orbital  separation of $14.13M$, when the initial separation was $20M$. 
The color scale is logarithmic and fixed to
emphasize the regions that saturate the interval $[10^{-7},10^{-1}]$.
The concentric dotted lines centered around the black hole and the origin 
indicate where the IZ-NZ BZ and the NZ-FZ BZ start and end respectively.
}
\label{fig:2D_R_IZNZ_GLO_comparison}
\end{figure*}

As a way of verifying that the second-order metric introduces fewer violations
to the Einstein equations than the first-order one, we have extended 
our analysis to a domain consisting of a two dimensional slice, the $z=0$ plane, 
including one of the black holes, its IZ, NZ, and BZ.
The slice is taken at a time $t=10756.60M$ (chosen because the BHs are again on the
$x$ axis) when the binary
separation is $r_{12}=14.13M$ after starting from $r_{12}=20M$ at $t=0M$.
Figure~\ref{fig:2D_R_IZNZ_GLO_comparison} makes it clear that the trends 
observed above remain, i.e., that 
the more accurate the global metric, the closer the Ricci scalar is to zero.
Thus, the smallness of the Ricci scalar computed with the second order metric 
is not due to plotting along the $x$ axis. In addition, this figure shows that the 
Ricci scalar decreases away from the black hole as expected from an 
approximation that asymptotically approaches Minkowski spacetime. 

\begin{figure*}[htb]
\begin{center}
\includegraphics[width=0.45\textwidth,clip=true]{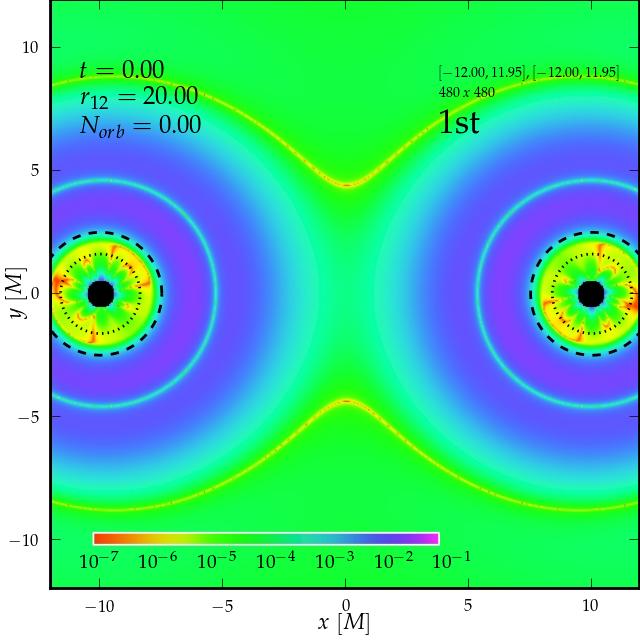}
\includegraphics[width=0.45\textwidth,clip=true]{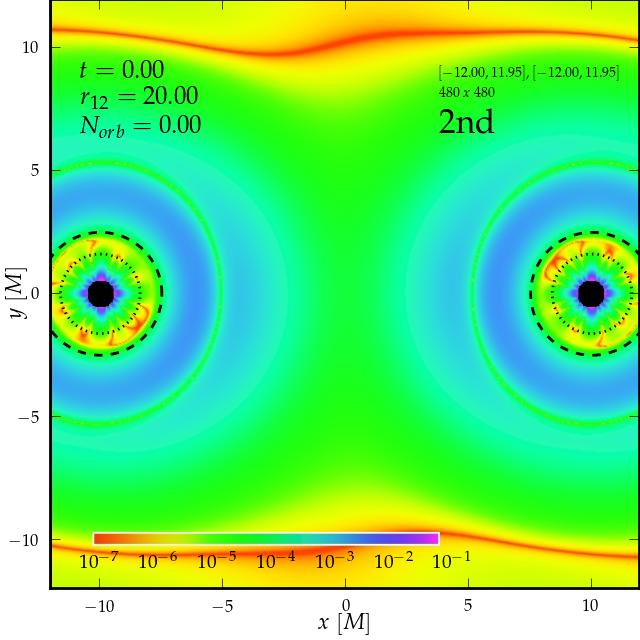}
\includegraphics[width=0.45\textwidth,clip=true]{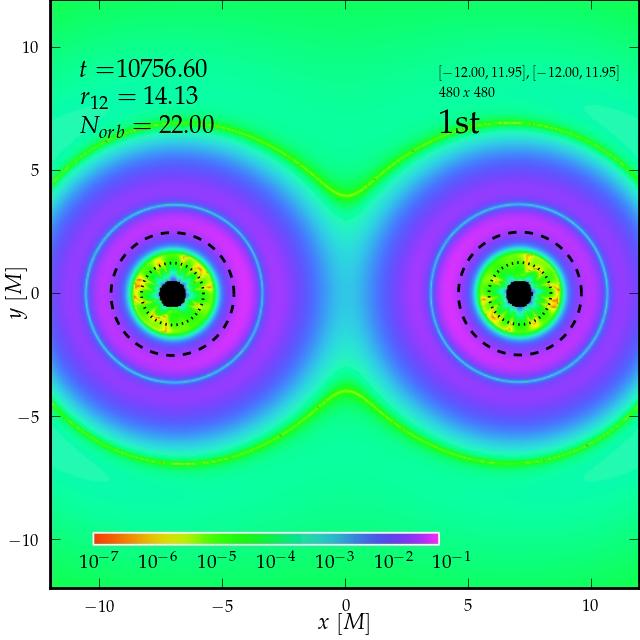}
\includegraphics[width=0.45\textwidth,clip=true]{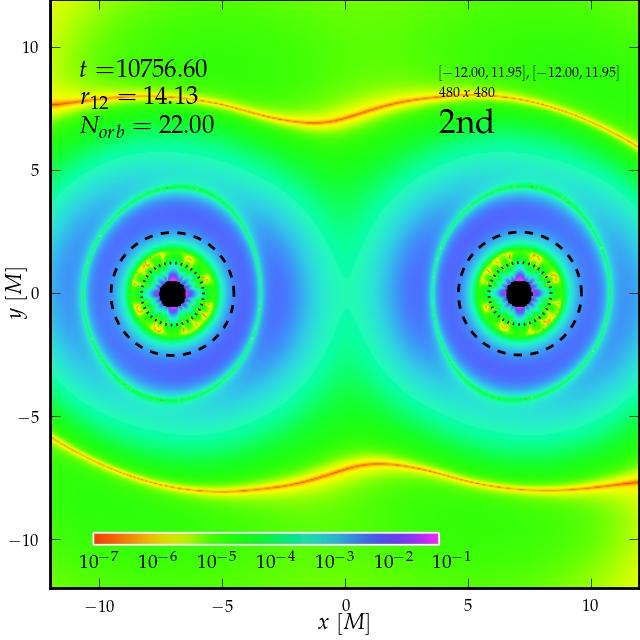}
\end{center}
\caption{(Color online) 
Absolute value of the Ricci scalar calculated with the
first-order (left) and the second-order (right) global metric at an
initial separation of $20 M$ (top) and after $10756.60 M$ of evolution, when the
separation is $14.13M$ (bottom). The color scale is logarithmic and fixed to
emphasize the regions that saturates the interval $[10^{-7},10^{-1}]$.
The concentric dotted and dashed lines around each black hole indicate where
the IZ-NZ BZ starts and where the ISCO would be located for an
individual black hole, respectively.
}
\label{fig:2D_R_b20_b14_GLO_comparison}
\end{figure*}

Finally, let us include both black holes in the domain and focus our attention on
the region between them and their immediate vicinity.
Figure~\ref{fig:2D_R_b20_b14_GLO_comparison} shows the Ricci scalar computed
with the first (left panels) and second-order metrics (right panels) at an initial separation of $20 M$ (top panels) and at $t=10756.60 M$ later (bottom panels), corresponding to a separation of $14.13 M$. This figure shows several interesting features. 
First, the second order metric is globally more accurate than the first-order metric as expected. 
Second, the accuracy of the approximations deteriorate with shrinking orbital separation, as expected for our construction. 
Third, the largest violations are in the IZ-NZ BZ, just outside where the ISCO would be if the black hole were isolated. These violations correspond to the ``humps'' observed originally in~\cite{Yunes:2005nn} when constructing initial data. 
Fourth, the Ricci scalar decreases towards the perpendicular bisector of the line joining both black holes. 

\section{CONCLUSIONS}
\label{sec:conclusions}

We have constructed a global, approximate spacetime metric that describes a
nonspinning, black hole binary system in a quasicircular inspiral trajectory.
This metric is built by asymptotically matching approximate metrics (a
perturbed Schwarzschild metric, a two-body PN metric, and a multipolar PM
metric) with a time-dependent transformation. We have evaluated the Ricci
scalar as a function of time to determine how accurately this global metric
satisfies the vacuum Einstein equations. We have found that indeed the metric
constructed is as accurate as expected, with dominant errors arising due to
uncontrolled remainders in the different approximations. 

Our immediate goal is to use this metric to study a variety of astrophysical
phenomena related to black hole binary spacetimes, such as the MHD evolution of
accretion disks around SMBH mergers starting from astrophysically relevant
initial conditions, the associated emission of electromagnetic
radiation from such mergers, and the formation of jets. Upcoming work in this
arena will be published in separate papers.

The specific spacetime work carried out here could be expanded in different directions. Perhaps
the obvious first step is to repeat the matching calculation from first
principles, without assuming $(t - t_{0})/r_{12} \ll 1$, as is done in (almost)
all previous GR matching studies. Doing so to highest order possible in the
matching may provide a more accurate transformation, which would thus reduce
the magnitude of the uncontrolled remainders in the present metric. Another
possible direction for future research would be to repeat this analysis for
spinning black hole binaries. Recently, Ref.~\cite{Gallouin:2012kb} used the
perturbed Kerr metric of~\cite{Yunes:2005ve} and the NZ PN metric
of~\cite{Faye:2006gx}  to compute an asymptotically matched metric on a short
slab of spatial hypersurfaces. This asymptotically matched metric is then
sufficient to construct a global metric, following the temporal-resummation
prescription developed here. 

Another possible avenue for future research would be to consider more generic
orbits. In this paper, we focused on quasicircular, nonspinning inspirals,
neglecting orbital eccentricity and precession. Recent work on eccentric
inspirals~\cite{Damour:1985,Gopakumar:2001dy,Yunes:2009yz} could be used to
extend the work
in~\cite{Yunes:2005nn,Yunes:2006iw,JohnsonMcDaniel:2009dq,Gallouin:2012kb} and
construct an asymptotically matched spacetime for binaries in eccentric orbits. 

Finally, it may be desirable to reconsider the construction of a global metric
in a coordinate system that is better adapted to numerical simulations. One
such set of coordinates are the ADM-TT ones, for example used in the PN work
of~\cite{Tichy:2002ec,Yunes:2006iw,Kelly:2007uc,Kelly:2009js,Mundim:2010hu}.
One could start by constructing an asymptotically matched metric using ADM-TT
coordinates in the NZ, as was done for example in~\cite{Yunes:2006iw}. Of course,
one would have to extend this work to next order in matching, which was 
accomplished in~\cite{JohnsonMcDaniel:2009dq} using harmonic coordinates. If this is done
without assuming an expansion about an initial spatial hypersurface, then the
resulting global metric would be valid for all times, while a BZ exists. 

\acknowledgments
M.~C., B.~M., H.~N., S.~N., and Y.~Z. gratefully acknowledge the National
Science Foundation (NSF) for financial support from Grants No. AST-1028087, 
No. PHY-1305730, No. PHY-1212426, No. PHY-1229173, No. PHY-1125915, No. PHY-0929114, No. PHY-0969855, No. PHY-0903782,
No. OCI-0832606, and No. DRL-1136221.  B.~M. was also supported in part by the DFG
Grant SFB/Transregio~7 and by the CompStar network, COST Action MP1304. H.~N.
would also like to acknowledge support by the Grant-in-Aid for Scientific
Research (No. 24103006).  N.~Y. acknowledges support from NSF Grant No. PHY-1114374,
NASA Grant No. NNX11AI49G, under subaward No. 00001944 and the NSF CAREER Award
No. PHY-1250636. S.N. was supported in part by the National Science Foundation
under Grant No. NSF PHY11-25915. Computational resources were provided by XSEDE
allocation TG-PHY060027N and by NewHorizons and BlueSky Clusters at Rochester
Institute of Technology, which were supported by NSF Grants No. PHY-0722703,
No. DMS-0820923, No. AST-1028087, and No. PHY-1229173.  The simulations were performed on
Stampede at TACC, Datura at AEI-Potsdam, and on LOEWE at CSC-Frankfurt.

\appendix

\section{TRANSITION FUNCTIONS}
\label{app:trans}

When constructing a global metric, appropriate transition functions must be
used to avoid introducing spurious error~\cite{Yunes:2006mx}. In this paper, we
used the following piecewise function in the BZs:
\begin{widetext}
\begin{eqnarray}
\label{trans-eq}
 f(r,\,r_0,\,w,\,q,\,s) =
 \left\{
     \begin{array}{lr}
       0 \,, & \hspace{-85mm} r \le r_0 \,, \\
       \frac{1}{2} \left\{ 1 +\tanh \left[\frac{s}{\pi} \left(\chi(r,\,r_0,\,w) - \frac{q^2}{\chi(r,\,r_0,\,w)} \right) \right] \right\} \,, 
           & \hspace{9mm}r_0 < r < r_0 + w \,, \\
       1 \,, & \hspace{-85mm} r \ge r_0 + w \,,     
     \end{array}
   \right.
\end{eqnarray}
\end{widetext}
where $\chi(r, r_0, w)=\tan[\pi(r - r_0)/(2w)]$, and $r_0$, $w$, $q$ and $s$
are parameters.
References~\cite{Yunes:2005nn,Yunes:2006iw,JohnsonMcDaniel:2009dq} discuss this
transition function in great detail. This function satisfies the Frankenstein
conditions of~\cite{Yunes:2006mx} for an appropriate set of parameters.

\begin{figure*}[htb]
\begin{center}
\includegraphics[width=0.45\textwidth,clip=true]{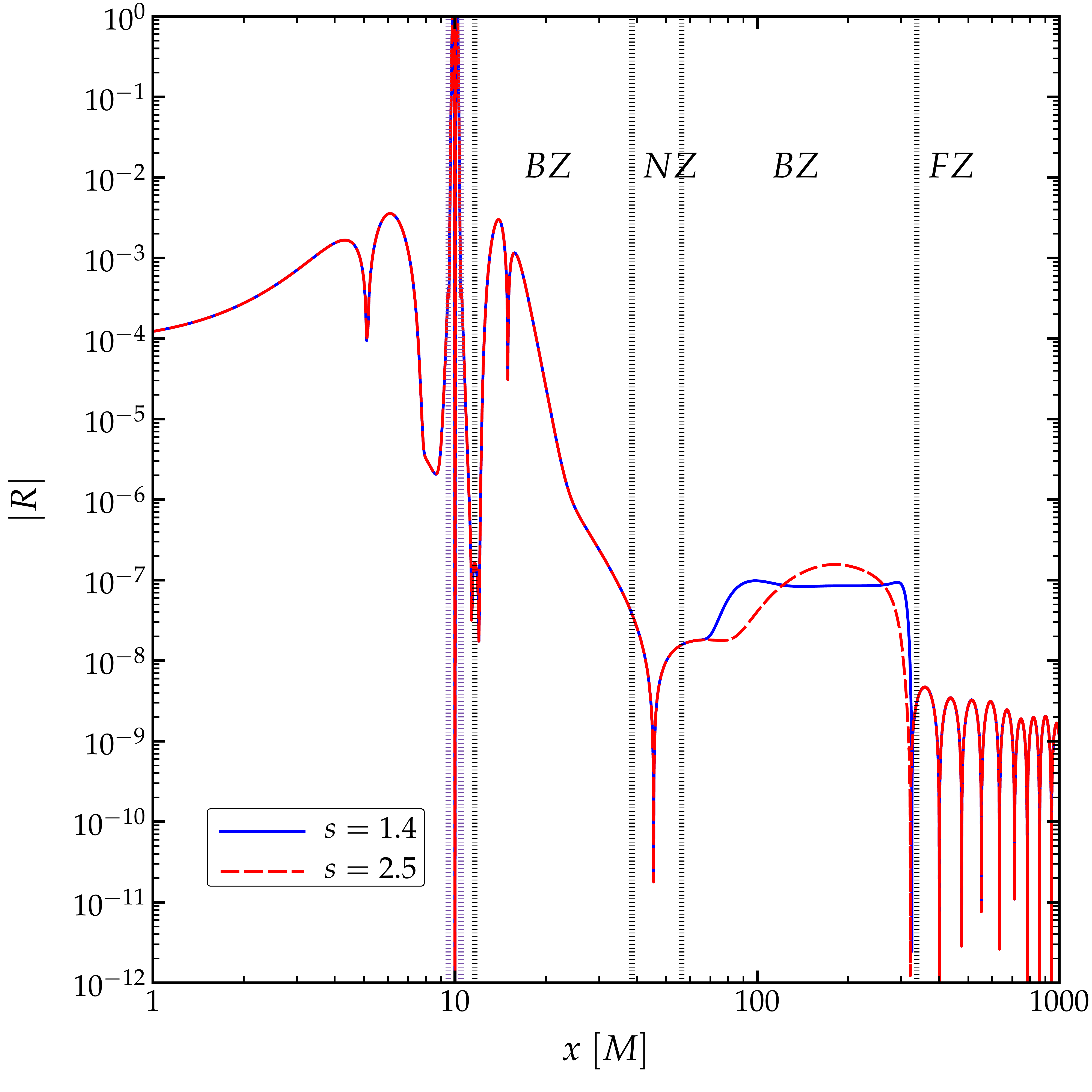}
\includegraphics[width=0.45\textwidth,clip=true]{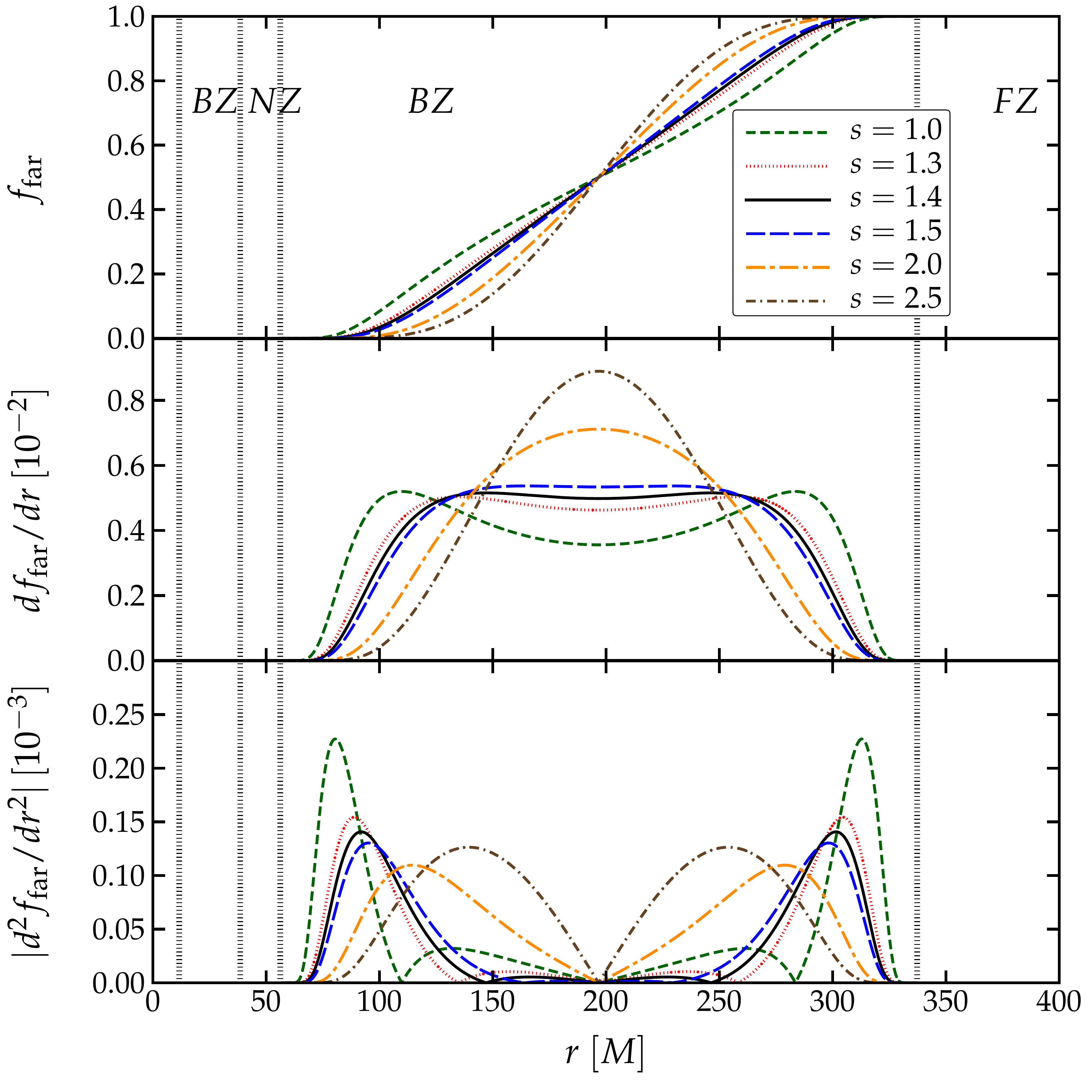}
\end{center}
\caption{(Color online) 
Left: L2 norm of the Ricci scalar along the $x$ axis computed with the second-order metric for different choices
of $s$ in the NZ-FZ BZ transition function. Observe that an appropriate choice of transition function parameter 
leads to a smaller ``bump'' in the Ricci scalar in the NZ-FZ BZ. Right: Behavior of $f_{\rm far}=f(r,\,\lambda/5,\,\lambda,\,1,\,s)$,
$df_{\rm far}/dr$ and $|d^2f_{\rm far}/dr^2|$ for various values of $s$.
Observe that the choice $s=1.4$ leads to a better behaved transition. 
}
\label{fig:transition_s2p5}
\end{figure*}

The transition functions in different BZs will have slightly different
parameters~\cite{JohnsonMcDaniel:2009dq}. We used the following throughout the paper 
(unless otherwise mentioned):
\begin{eqnarray}
 f_{\rm near} &=& f(x,\, 2.2 m_2 - m_1 r_{12}/M,\, r_{12} - 2.2 M,\, 1,\, 1.4) \,,
 \nonumber \\
 f_{{\rm inner},A} &=& f(r_A,\, 0.256 r_A^{\rm T},\, 3.17 (M^2 r_{12}^5)^{1/7},\, 0.2,\, r_{12}/M) \,.
\cr &&
\end{eqnarray}
The transition radius $r_A^{\rm T} = (m_A^3 r_{12}^5 / M)^{1/7}$ was derived by
requiring that the uncontrolled  remainders of the approximations in the IZ and
NZ be comparable. The transition function in the NZ-FZ BZ was also
given in~\cite{JohnsonMcDaniel:2009dq} by Eq.~\eqref{trans-eq} with the
parameters $r_{0} = \lambda/5$, $w = \lambda$, $q = 1$ and $s = 2.5$, where
$\lambda = \pi\sqrt{r_{12}^3/M}$ in the Newtonian limit. We found that this
transition function leads to a rather large ``bump'' in the Ricci scalar $|R|$
around $x/M\approx200$, where the transition between the NZ and FZ metrics
occurs as shown in the left panel of Fig.~\ref{fig:transition_s2p5}.

In order to remove this undesired behavior, let us study this transition
function as a function of $s$ at $r_{12}/M = 20$. The right panel of
Fig.~\ref{fig:transition_s2p5}
shows $f_{\rm far}$ as well as its derivatives $df_{\rm far}/dr$ and
$|d^2f_{\rm far}/dr^2|$ for various values of $s$. This figure suggests that
any value of $s$ between $1.3$ and $1.5$ would lead to a better 
behaved NZ-FZ transition, so we here choose $s = 1.4$. 
We have verified that with this choice of $s$, the Frankenstein
theorems of~\cite{Yunes:2006mx} are still satisfied.

\section{SOME DETAILS TO CONSTRUCT THE GLOBAL METRIC}
\label{app:equations}

In this appendix, we present in more detail the equations used to construct the global metric.
In order not to duplicate equations in previous studies, let us first describe which equations
we have used from previously published papers. For ease of reading, Eq.~(n) from 
Ref.~\cite{JohnsonMcDaniel:2009dq} will be denoted as Eq.~(JM-n).
For the NZ, the PN metric is given in Eq.~(JM-4.1)
(see also Eq.~(7.2) of Ref.~\cite{Blanchet:1998vx} for completeness).
Orbital information required to calculate the metric is taken from
Ref.~\cite{Blanchet:2006zz}. For example, the phase evolution for quasicircular, 
nonspinning inspirals is given in Eq.~(234) of Ref.~\cite{Blanchet:2006zz}.
The FZ metric is obtained from Eq.~(JM-6.3) with Eq.~(JM-6.1),
which is determined by the source multipoles in Eq.~(JM-6.2).
The IZ metric is given by Eq.~(JM-3.2) with Eq.~(JM-3.3).
This metric is prescribed by multipole tidal fields
in IZ coordinates $\{T,X,Y,Z\}$.
The multipole tidal fields are given in Eq.~(JM-B.1).

The temporally resummed coordinate transformation found in this paper 
(i.e., the mapping between the IZ coordinates $\{T,X,Y,Z\}$ and the PN harmonic 
coordinates $\{t,x,y,z\}$) to second order in the matching
is explicitly 
\begin{widetext}
\begin{align}
T &= 
t-{\frac {{m_2}\,{\tilde{y}_c}}{\sqrt {{r_{12}}}\sqrt {M}}}
+\frac{5}{384} \frac{(m_2+2 M)(r_{12}^3-r_{12}(t_0)^3)}{M^2 m_1}
+\biggl[ -\frac{1}{2}\,{\frac {{m_2}\,{\tilde{y}}\,\sqrt {M}{\tilde{x}}\,{{\tilde{r}}}^{2}}{{{r_{12}}}^{9/2}}}
+\frac{1}{2}\,{\frac {{m_2}\,{\tilde{y}_c}\,\sqrt {M}{{\tilde{r}}}^{2}}{{{r_{12}}}^{7/2}}}
+\frac{5}{2}\,{\frac {{m_2}\,{\tilde{y}}\,{{\tilde{x}}}^{3}\sqrt {M}}{{{r_{12}}}^{9/2}}}
\cr & \quad
-2\,{\frac {{{\tilde{x}}}^{2}\sqrt {M}{\tilde{y}}\,{m_2}}{{{r_{12}}}^{7/2}}}
-{\frac {{m_2}\,{\tilde{y}_c}\,{\tilde{x}_c}\, \left( {m_2}-2\,M \right) }{\sqrt {M}{{r_{12}}}^{5/2}}}
+\frac{1}{2}\,{\frac {{m_2}\,{\tilde{y}_c}\, \left( -5\,M+{m_2} \right) }{\sqrt {M}{{r_{12}}}^{3/2}}} \biggr]
\cr & \quad
+\biggl[ -4\,{\frac { \left( {m_2}-M \right) ^{2}}{{\tilde{r}}}}
+2\,{\frac {{m_2}\,\sqrt {M}{\tilde{x}_c}\, \left( -2\,{{\tilde{y}_c}}^{2}
+{{\tilde{x}_c}}^{2} \right) \sin{\phi} 
\cos{\phi} }{{r_{12}}^{7/2}}}
\cr & \qquad
+{\frac {5}{1024}}
\,{\frac { \left( 4\,{M}^{3}-4\,{m_2}\,{M}^{2}-3\,{{m_2}}^{3} \right)  
(r_{12}^2-r_{12}(t_0)^2)}{{m_1}\,{M}^{3}}} \biggr] 
\cr & \quad
+\biggl[ \frac{1}{12}\,{\frac {{m_2}\,{\tilde{y}_c}\, \left( 6\,{{m_2}}^{2}
+7\,{m_2}\,M+5\,{M}^{2} \right) {{\tilde{r}}}^{2}}{\sqrt {M}{r_{12}}^{9/2}}}
-\frac{14}{3}\,{\frac {{m_2}\,{\tilde{y}_c}\,\sqrt {M}{\tilde{x}_c}\, 
\left( {m_2}-M \right) {\tilde{r}}}{{r_{12}}^{9/2}}}
\cr & \qquad
-{\frac {{m_2}\,{\tilde{y}_c}
\,\sqrt {M} \left( {m_2}-M \right) {\tilde{r}}}{{r_{12}}^{7/2}}}
+\frac{1}{4}\,{\frac {{m_2}\,{\tilde{y}_c}\, \left( M+{m_2} \right) {{\tilde{r}}}^{2} 
\left( \sin{\phi}  \right) ^{2} 
\left( \cos{\phi}  \right) ^{2}}{\sqrt {M}{r_{12}}^{7/2}}}
\cr & \qquad
-\frac{1}{3}\,{\frac {{m_2}\,{\tilde{y}_c}\, \left( 6\,{{\tilde{y}_c}}^{2}M{m_2}+9\,{{\tilde{x}_c}}^{2}
{{m_2}}^{2}-3\,{{\tilde{y}_c}}^{2}{{m_2}}^{2}-{{\tilde{y}_c}}^{2}{M}^{2}-7\,M{{\tilde{x}_c}}^{2}
{m_2}+10\,{{\tilde{x}_c}}^{2}{M}^{2} \right) }{\sqrt {M}{r_{12}}^{9/2}}}
\cr & \qquad
+\frac{1}{2}\,{\frac {{m_2}\,{\tilde{y}_c}\,{\tilde{x}_c}\, \left( 2\,{M}^{2}+3\,{{m_2}}^{2}
-3\,{m_2}\,M \right) }{\sqrt {M}{r_{12}}^{7/2}}}
+\frac{1}{12}\,{\frac {{{m_2}}^{2}{\tilde{y}_c}
\, \left( 17\,{m_2}\,{M}^{2}+30\,{{m_2}}^{3}
-17\,{{m_2}}^{2}M+12\,{M}^{3} \right) }{{M}^{5/2}{r_{12}}^{5/2}}}
\cr & \qquad
-\frac{1}{4}\,{\frac {{m_2}\,{\tilde{y}_c}\, \left( 56\,M{{\tilde{x}_c}}^{2}
-16\,{{\tilde{y}_c}}^{2}M+13\,{m_2}\,{{\tilde{x}_c}}^{2} \right)  
\left( \sin{\phi}  \right) ^{2} 
\left( \cos{\phi}  \right) ^{2}}{\sqrt {M}{r_{12}}^{7/2}}}
\cr & \qquad
+\frac{1}{2}\,{\frac {{m_2}\,{\tilde{y}_c}\,{\tilde{x}_c}\, \left( -19\,{m_2}\,M
+3\,{{m_2}}^{2}+8\,{M}^{2} \right)  \left( \sin{\phi}  \right) ^{2} 
\left( \cos{\phi}  \right) ^{2}}{{M}^{3/2}{r_{12}}^{5/2}}}
+\frac{1}{24}\,{\frac {{m_2}\,{\tilde{y}_c}\, \left( \sin{\phi}  \right) ^{4} 
\left( \cos{\phi}  \right) ^{4}}{\sqrt {r_{12}}\sqrt {M}}} \biggr]
\,,
\cr
X &=
{\tilde{x}}
+\left[ -\frac{1}{2}\,{\frac {{{m_2}}^{2}\sin{\phi}\,{\tilde{y}_c}}{M{r_{12}}}}
+{\tilde{x}}\, \left( {\dot{A}}-\frac{1}{2}\,{\frac {{{m_2}}^{2}}{M{r_{12}}}} \right) 
+\frac{1}{2}\,{\frac {{m_2}\, \left( -2\,{\tilde{x}}\,{\tilde{x}_c}+\cos{\phi}\,{{\tilde{r}}}^{2} \right) }{{{r_{12}}}^{2}}} \right] 
-{\tilde{y}}\,{R^z} 
\cr & \quad
+\biggl[ -\frac{1}{8}\,{\frac {{m_2}\, \left( -M+2\,{m_2} \right) {{\tilde{r}}}^{4}}{{r_{12}}^{5}}}
+\frac{1}{4}\,{\frac {{m_2}\, \left( -{{\tilde{y}_c}}^{2}M-7\,M{{\tilde{x}_c}}^{2}
+6\,{m_2}\,{{\tilde{x}_c}}^{2} \right) {{\tilde{r}}}^{2}}{{r_{12}}^{5}}}
-\frac{1}{2}\,{\frac {{m_2}\, \left( {m_2}-M \right) {{\tilde{r}}}^{3}}{{r_{12}}^{4}}}
\cr & \qquad
-{\frac {{m_2}\,{\tilde{x}_c}\, \left( {m_2}-M \right) {{\tilde{r}}}^{2}}{{r_{12}}^{4}}}
+3\,{\frac {{m_2}\,{{\tilde{x}_c}}^{2} \left( {m_2}-M \right) {\tilde{r}}}{{r_{12}}^{4}}}
-\frac{5}{2}\,{\frac {{m_2}\,{{\tilde{x}_c}}^{4} \left( {m_2}-M \right) }{{r_{12}}^{4}{\tilde{r}}}}
+\frac{1}{2}\,{\frac {{m_2}\, \left( -3\,M+4\,{m_2} \right) {{\tilde{r}}}^{2}}{{r_{12}}^{3}}}
\cr & \qquad
-3\,{\frac {{m_2}\,{\tilde{x}_c}\, \left( {m_2}-M \right) {\tilde{r}}}{{r_{12}}^{3}}}
+3\,{\frac {{m_2}\,{{\tilde{x}_c}}^{3} \left( {m_2}-M \right) }{{r_{12}}^{3}{\tilde{r}}}}
-2\,{\frac {\cos{\phi} {m_2}\, \left( {m_2}-M \right) {\tilde{r}}}{{r_{12}}^{2}}}
\cr & \qquad
-\frac{1}{4}\,{\frac {{m_2}\,{{\tilde{x}_c}}^{2} \left( -7\,M{{\tilde{x}_c}}^{2}+5\,{m_2}\,{{\tilde{x}_c}}^{2}
-11\,{{\tilde{y}_c}}^{2}M \right) }{{r_{12}}^{5}}}+{\frac {{m_2}\,{\tilde{x}_c}\, 
\left( {m_2}\,{{\tilde{x}_c}}^{2}+{{\tilde{y}_c}}^{2}{m_2}-M{{\tilde{x}_c}}^{2}
-2\,{{\tilde{y}_c}}^{2}M \right) }{{r_{12}}^{4}}}
\cr & \qquad
+ \left( -{m_2}\, 
\left( 3\,{m_2}-2\,M \right) {{\tilde{x}_c}}^{2}+{\frac {{m_2}\, \left( {m_2}-M \right) ^{2}
{{\tilde{y}_c}}^{2}}{M}} \right) \frac{1}{r_{12}^{3}}
+\frac{1}{2}\,{\frac {{m_2}\, \left( -7\,M+8\,{m_2} \right)  
\left( {\tilde{x}_c}\,\cos{\phi} +{\tilde{y}_c}\,\sin{\phi}  \right) }{{r_{12}}^{2}}} \biggr]
\cr & \quad
+\biggl[ -3\,{\frac {{m_2}\,{\tilde{y}_c}\,{\tilde{x}_c}\, \left( {m_2}-M \right) \sin{\phi} 
\cos{\phi} {{\tilde{r}}}^{2}}{{r_{12}}^{5}}}
-{\frac {{m_2}\,{\tilde{y}_c}\, \left( {{m_2}}^{2}+{M}^{2}-{m_2}\,M \right) 
\sin{\phi} 
\cos{\phi} {{\tilde{r}}}^{2}}{M{r_{12}}^{4}}}
\cr & \qquad
-6\,{\frac {{m_2}\,{\tilde{y}_c}\,{\tilde{x}_c}\, \left( {m_2}-M \right) 
\sin{\phi} \cos{\phi} {\tilde{r}}}{{r_{12}}^{4}}}
+10\,{\frac {{m_2}\,{{\tilde{x}_c}}^{3}{\tilde{y}_c}\, \left( {m_2}-M \right) 
\sin{\phi} \cos{\phi} }
{{r_{12}}^{4}{\tilde{r}}}}-3\,{\frac {{m_2}\,{{\tilde{x}_c}}^{2}{\tilde{y}_c}\, 
\left( {m_2}-M \right) \sin{\phi} 
\cos{\phi} }{{r_{12}}^{3}{\tilde{r}}}}
\cr & \qquad
+\frac{1}{2}\,{\frac {{m_2}\,{\tilde{y}_c}\,{\tilde{x}_c}\, \left( 10\,{m_2}\,{{\tilde{x}_c}}^{2}
-11\,{{\tilde{y}_c}}^{2}M-3\,M{{\tilde{x}_c}}^{2} \right) \sin{\phi} 
\cos{\phi} }{{r_{12}}^{5}}}
\cr & \qquad
+ \left( {\frac {{m_2}\, \left( -{M}^{2}+3\,{{m_2}}^{2}+{m_2}\,M \right) 
{\tilde{y}_c}\,{{\tilde{x}_c}}^{2}}{M}}
-\frac{1}{2}\,{m_2}\, 
\left( -4\,M+3\,{m_2} \right) {{\tilde{y}_c}}^{3} \right) 
\frac{\sin{\phi} \cos{\phi}}{{r_{12}}^{4}}
\cr & \qquad
+{\frac {{m_2}\,{\tilde{y}_c}\,{\tilde{x}_c}\, \left( {{m_2}}^{2}-4\,{m_2}\,M+2\,{M}^{2} \right) 
\sin{\phi} \cos{\phi} }
{M{r_{12}}^{3}}}
+\frac{1}{4}\,{\frac {{m_2}\,{\tilde{y}_c}\, \left( {{m_2}}^{3}
-5\,{{m_2}}^{2}M-4\,{M}^{3}-{m_2}\,{M}^{2} \right) \sin{\phi} 
\cos{\phi} }{{M}^{2}{r_{12}}^{2}}}
\cr & \qquad
-\frac{1}{6}\,{\frac {{m_2}\,{\tilde{y}_c}\,{\tilde{x}_c}\, \left( \sin{\phi}  \right) ^{3} 
\left( \cos{\phi}  \right) ^{3}}{{r_{12}}^{2}}}
-\frac{1}{6}\,{\frac {{m_2}\,{\tilde{y}_c}\, \left( -8\,M+5\,{m_2} \right)  
\left( \sin{\phi}  \right) ^{3} 
\left( \cos{\phi}  \right) ^{3}}{Mr_{12}}} \biggr]
\,,
\cr
Y &= 
{\tilde{y}}
+\left[ \frac{1}{2}\,{\frac {{{m_2}}^{2}\cos{\phi}\,{\tilde{y}_c}}{M{r_{12}}}}
+{\tilde{y}}\, \left( {\dot{A}}-\frac{1}{2}\,{\frac {{{m_2}}^{2}}{M{r_{12}}}} \right) 
+\frac{1}{2}\,{\frac {{m_2}\, \left( -2\,{\tilde{y}}\,{\tilde{x}_c}+\sin{\phi}\,{{\tilde{r}}}^{2} \right) }{{{r_{12}}}^{2}}} \right]
+{\tilde{x}}\,{R^z}
\cr & \quad
+\biggl[ \frac{3}{4}\,{\frac {{\tilde{x}_c}\,{\tilde{y}_c}\,{m_2}\, \left( -3\,{m_2}+M \right) {{\tilde{r}}}^{2}}{{r_{12}}^{5}}}
+\frac{1}{2}\,{\frac {{\tilde{x}_c}\,{\tilde{y}_c}\,{m_2}\, \left( {m_2}-M \right) {\tilde{r}}}{{r_{12}}^{4}}}
-\frac{5}{2}\,{\frac {{\tilde{y}}\,{m_2}\,{{\tilde{x}}}^{3} \left( {m_2}-M \right) }{{r_{12}}^{4}{\tilde{r}}}}
+3\,{\frac {{\tilde{y}_c}\,{m_2}\,{{\tilde{x}_c}}^{2} \left( {m_2}-M \right) }{{r_{12}}^{3}{\tilde{r}}}}
\cr & \qquad
-\frac{1}{4}\,{\frac {{\tilde{x}_c}\,{\tilde{y}_c}\,{m_2}\, \left( -21\,{m_2}\,{{\tilde{x}_c}}^{2}
-6\,{{\tilde{y}_c}}^{2}M+13\,M{{\tilde{x}_c}}^{2} \right) }{{r_{12}}^{5}}}
+{\frac {{\tilde{y}_c}\,{m_2}\, \left( 3\,{{\tilde{x}_c}}^{2}-{{\tilde{y}_c}}^{2} \right)  
\left( M-{m_2} \right) }{{r_{12}}^{4}}}
-\frac{1}{2}\,{\frac {{\tilde{x}_c}\,{\tilde{y}_c}\,{m_2}\, 
\left( 2\,M-{m_2} \right) ^{2}}{M{r_{12}}^{3}}}
\cr & \qquad
+\frac{1}{8}\,{\frac {{m_2}\, \left( {\tilde{y}_c}\,\cos{\phi} 
-{\tilde{x}_c}\,\sin{\phi}  \right)  
\left( -16\,{m_2}\,{M}^{2}+8\,{{m_2}}^{2}M-{{m_2}}^{3}+20\,{M}^{3} \right) }{{M}^{2}{r_{12}}^{2}}}
-2\,{\frac {\sin{\phi} {m_2}\, \left( {m_2}-M \right) 
{\tilde{r}}\,\cos{\phi} }{{r_{12}}^{2}}} \biggr]
\cr & \quad
+\biggl[ \frac{3}{4}\,{\frac {{m_2}\, \left( {\tilde{y}_c}-{\tilde{x}_c} \right)  
\left( {\tilde{y}_c}+{\tilde{x}_c} \right)  \left( -M+3\,{m_2} \right) 
\sin{\phi} \cos{\phi} {{\tilde{r}}}^{2}}{{r_{12}}^{5}}}
-\frac{3}{4}\,{\frac {{\tilde{x}_c}\,{{m_2}}^{2} \left( -M+3\,{m_2} \right) 
\sin{\phi} \cos{\phi} {{\tilde{r}}}^{2}}{M{r_{12}}^{4}}}
\cr & \qquad
-\frac{1}{2}\,{\frac {{m_2}\, \left( {\tilde{y}_c}-{\tilde{x}_c} \right)  
\left( {\tilde{y}_c}+{\tilde{x}_c} \right)  \left( {m_2}-M \right) 
\sin{\phi} \cos{\phi} {\tilde{r}}}
{{r_{12}}^{4}}}
+\frac{1}{4}\,{\frac {{m_2}\, \left( 4\,{m_2}\,M-6\,{M}^{2}+{{m_2}}^{2} \right) 
\sin{\phi} \cos{\phi} {{\tilde{r}}}^{2}}{M{r_{12}}^{3}}}
\cr & \qquad
-3\,{\frac {{m_2}\,{\tilde{x}_c}\, \left( {m_2}-M \right) \sin{\phi} 
\cos{\phi} {\tilde{r}}}{{r_{12}}^{3}}}
+3\,{\frac {{m_2}\,{{\tilde{x}_c}}^{3} \left( {m_2}-M \right) 
\sin{\phi} \cos{\phi} }{{r_{12}}^{3}{\tilde{r}}}}
-8\,{\frac {{m_2}\,{\tilde{x}_c}\,\sqrt {M} \left( {m_2}-M \right) }{{r_{12}}^{5/2}}}
\cr & \qquad
-\frac{1}{4}\,{\frac {{m_2}\, \left( 13\,{{\tilde{x}_c}}^{4}M-57\,M{{\tilde{x}_c}}^{2}{{\tilde{y}_c}}^{2}
+63\,{m_2}\,{{\tilde{y}_c}}^{2}{{\tilde{x}_c}}^{2}+6\,M{{\tilde{y}_c}}^{4}-21\,{m_2}\,{{\tilde{x}_c}}^{4} \right) 
\sin{\phi} \cos{\phi} }{{r_{12}}^{5}}}
\cr & \qquad
-\frac{1}{4}\,{\frac {{m_2}\,{\tilde{x}_c}\, \left( -60\,{{\tilde{y}_c}}^{2}M{m_2}+25\,M{{\tilde{x}_c}}^{2}{m_2}
+36\,{{\tilde{y}_c}}^{2}{M}^{2}-12\,{{\tilde{x}_c}}^{2}{M}^{2}-21\,{{\tilde{x}_c}}^{2}{{m_2}}^{2}
+18\,{{\tilde{y}_c}}^{2}{{m_2}}^{2} \right) \sin{\phi} 
\cos{\phi} }{M{r_{12}}^{4}}}
\cr & \qquad
+\frac{1}{2}\,{\frac {{m_2}\, \left( -2\,M{{\tilde{x}_c}}^{2}{m_2}+2\,{{\tilde{y}_c}}^{2}{M}^{2}
+4\,{{\tilde{x}_c}}^{2}{M}^{2}+2\,{{\tilde{y}_c}}^{2}{{m_2}}^{2}-{{\tilde{x}_c}}^{2}{{m_2}}^{2}
-4\,{{\tilde{y}_c}}^{2}M{m_2} \right) \sin{\phi} 
\cos{\phi} }{M{r_{12}}^{3}}}
\cr & \qquad
-\frac{1}{4}\,{\frac {{m_2}\,{\tilde{x}_c}\, \left( 2\,{{m_2}}^{3}-4\,{M}^{3}-9\,{{m_2}}^{2}M
-{m_2}\,{M}^{2} \right) \sin{\phi} 
\cos{\phi} }{{M}^{2}{r_{12}}^{2}}}
\cr & \qquad
+\frac{1}{24}\,{\frac {{{m_2}}^{2} \left( -16\,{{m_2}}^{2}M+40\,{m_2}\,{M}^{2}
+24\,{M}^{3}+51\,{{m_2}}^{3} \right) \sin{\phi} 
\cos{\phi} }{{M}^{3}r_{12}}}
\cr & \qquad
-\frac{1}{12}\,{\frac {{m_2}\, 
\left( 2\,{{m_2}}^{2}{r_{12}}^{2}-16\,{m_2}\,M{r_{12}}^{2}+16\,{M}^{2}r_{12}{\tilde{x}_c}
-2\,{\tilde{x}_c}\,M{m_2}\,r_{12}+2\,{{\tilde{y}_c}}^{2}{M}^{2}-{M}^{2}{{\tilde{r}}}^{2} \right)  
\left( \sin{\phi}  \right) ^{3} 
\left( \cos{\phi}  \right) ^{3}}{{M}^{2}{r_{12}}^{2}}}
\cr & \qquad
+\frac{1}{6}\,{\frac {{{m_2}}^{2} \left( -8\,M+{m_2} \right)  
\left( \sin{\phi}  \right) ^{3} 
\left( \cos{\phi}  \right) ^{3}}{{M}^{2}}}
+{\frac {1}{120}}\,{\frac {{m_2}\, \left( \sin{\phi}  \right) ^{5} 
\left( \cos{\phi}  \right) ^{5}r_{12}}{M}} \biggr]
\,,
\cr
Z &=
z
+ \left[ z \left( {\dot{A}}-\frac{1}{2}\,{\frac {{{m_2}}^{2}}{M{r_{12}}}} \right) 
-{\frac {{m_2}\,z{\tilde{x}_c}}{{{r_{12}}}^{2}}} \right] 
\cr & \quad
+\biggl[ \frac{1}{4}\,{\frac {{m_2}\,z{\tilde{x}_c}\, \left( -9\,{m_2}
+M \right) {{\tilde{r}}}^{2}}{{r_{12}}^{5}}}
+\frac{1}{2}\,{\frac {z{{m_2}}^{2}{{\tilde{r}}}^{2}}{{r_{12}}^{4}}}
-\frac{1}{2}\,{\frac {{m_2}\,z{\tilde{x}_c}\, \left( M-{m_2} \right) {\tilde{r}}}{{r_{12}}^{4}}}
+\frac{5}{2}\,{\frac {{m_2}\,z{{\tilde{x}_c}}^{3} \left( M-{m_2} \right) }{{r_{12}}^{4}{\tilde{r}}}}
\cr & \qquad
-3\,{\frac {{m_2}\,z{{\tilde{x}_c}}^{2} \left( M-{m_2} \right) }
{{r_{12}}^{3}{\tilde{r}}}}
-\frac{3}{4}\,{\frac {{m_2}\,z{\tilde{x}_c}\, 
\left( -7\,{m_2}\,{{\tilde{x}_c}}^{2}-2\,{{\tilde{y}_c}}^{2}M+M{{\tilde{x}_c}}^{2} \right) }{{r_{12}}^{5}}}
\cr & \qquad
+{\frac {{m_2}\,z \left( {{\tilde{y}_c}}^{2}{m_2}-{{\tilde{y}_c}}^{2}M
-3\,{m_2}\,{{\tilde{x}_c}}^{2}+M{{\tilde{x}_c}}^{2} \right) }{{r_{12}}^{4}}}
+\frac{1}{2}\,{\frac {{m_2}\,z \left( M-{m_2} \right) }{{r_{12}}^{2}}} \biggr] 
\cr & \quad
+\biggl[ \frac{1}{4}\,{\frac {{\tilde{y}_c}\,{m_2}\,z \left( 9\,{m_2}-M \right) 
\sin{\phi} \cos{\phi} 
{{\tilde{r}}}^{2}}{{r_{12}}^{5}}}
-\frac{1}{2}\,{\frac {{\tilde{y}_c}\,{m_2}\,z \left( {m_2}-M \right) 
\sin{\phi} \cos{\phi} {\tilde{r}}}{{r_{12}}^{4}}}
\cr & \qquad
+\frac{15}{2}\,{\frac {{\tilde{y}_c}\,{m_2}\,{{\tilde{x}_c}}^{2}z 
\left( {m_2}-M \right) \sin{\phi} \cos{\phi} }{{r_{12}}^{4}{\tilde{r}}}}
-\frac{3}{4}\,{\frac {{\tilde{y}_c}\,{m_2}\,z \left( -7\,M{{\tilde{x}_c}}^{2}+2\,{{\tilde{y}_c}}^{2}M
+21\,{m_2}\,{{\tilde{x}_c}}^{2} \right) \sin{\phi} 
\cos{\phi} }{{r_{12}}^{5}}}
\cr & \qquad
-\frac{1}{2}\,{\frac {{\tilde{x}_c}\,{\tilde{y}_c}\,{m_2}\,z \left( 8\,{M}^{2}-23\,{m_2}\,M+9\,{{m_2}}^{2} \right) 
\sin{\phi} \cos{\phi} }{M{r_{12}}^{4}}}
-\frac{1}{6}\,{\frac {{\tilde{y}_c}\,{m_2}\,z \left( \sin{\phi}  \right) ^{3} 
\left( \cos{\phi}  \right) ^{3}}{{r_{12}}^{2}}} \biggr]
\,.
\end{align}
\end{widetext}
The second-order transformed IZ metric under the PN harmonic coordinates
is calculated by using the above equations.

\bibliography{./bhm_references}

\begin{thebibliography}{56}%
\makeatletter
\providecommand \@ifxundefined [1]{%
 \@ifx{#1\undefined}
}%
\providecommand \@ifnum [1]{%
 \ifnum #1\expandafter \@firstoftwo
 \else \expandafter \@secondoftwo
 \fi
}%
\providecommand \@ifx [1]{%
 \ifx #1\expandafter \@firstoftwo
 \else \expandafter \@secondoftwo
 \fi
}%
\providecommand \natexlab [1]{#1}%
\providecommand \enquote  [1]{``#1''}%
\providecommand \bibnamefont  [1]{#1}%
\providecommand \bibfnamefont [1]{#1}%
\providecommand \citenamefont [1]{#1}%
\providecommand \href@noop [0]{\@secondoftwo}%
\providecommand \href [0]{\begingroup \@sanitize@url \@href}%
\providecommand \@href[1]{\@@startlink{#1}\@@href}%
\providecommand \@@href[1]{\endgroup#1\@@endlink}%
\providecommand \@sanitize@url [0]{\catcode `\\12\catcode `\$12\catcode
  `\&12\catcode `\#12\catcode `\^12\catcode `\_12\catcode `\%12\relax}%
\providecommand \@@startlink[1]{}%
\providecommand \@@endlink[0]{}%
\providecommand \url  [0]{\begingroup\@sanitize@url \@url }%
\providecommand \@url [1]{\endgroup\@href {#1}{\urlprefix }}%
\providecommand \urlprefix  [0]{URL }%
\providecommand \Eprint [0]{\href }%
\providecommand \doibase [0]{http://dx.doi.org/}%
\providecommand \selectlanguage [0]{\@gobble}%
\providecommand \bibinfo  [0]{\@secondoftwo}%
\providecommand \bibfield  [0]{\@secondoftwo}%
\providecommand \translation [1]{[#1]}%
\providecommand \BibitemOpen [0]{}%
\providecommand \bibitemStop [0]{}%
\providecommand \bibitemNoStop [0]{.\EOS\space}%
\providecommand \EOS [0]{\spacefactor3000\relax}%
\providecommand \BibitemShut  [1]{\csname bibitem#1\endcsname}%
\let\auto@bib@innerbib\@empty
\bibitem [{\citenamefont {{Kaiser}}\ \emph {et~al.}(2010)\citenamefont
  {{Kaiser}}, \citenamefont {{Burgett}}, \citenamefont {{Chambers}},
  \citenamefont {{Denneau}}, \citenamefont {{Heasley}}, \citenamefont
  {{Jedicke}}, \citenamefont {{Magnier}}, \citenamefont {{Morgan}},
  \citenamefont {{Onaka}},\ and\ \citenamefont
  {{Tonry}}}]{2010SPIE.7733E..12K}%
  \BibitemOpen
  \bibfield  {author} {\bibinfo {author} {\bibfnamefont {N.}~\bibnamefont
  {{Kaiser}}}, \bibinfo {author} {\bibfnamefont {W.}~\bibnamefont {{Burgett}}},
  \bibinfo {author} {\bibfnamefont {K.}~\bibnamefont {{Chambers}}}, \bibinfo
  {author} {\bibfnamefont {L.}~\bibnamefont {{Denneau}}}, \bibinfo {author}
  {\bibfnamefont {J.}~\bibnamefont {{Heasley}}}, \bibinfo {author}
  {\bibfnamefont {R.}~\bibnamefont {{Jedicke}}}, \bibinfo {author}
  {\bibfnamefont {E.}~\bibnamefont {{Magnier}}}, \bibinfo {author}
  {\bibfnamefont {J.}~\bibnamefont {{Morgan}}}, \bibinfo {author}
  {\bibfnamefont {P.}~\bibnamefont {{Onaka}}}, \ and\ \bibinfo {author}
  {\bibfnamefont {J.}~\bibnamefont {{Tonry}}},\ }in\ \href {\doibase
  10.1117/12.859188} {\emph {\bibinfo {booktitle} {Society of Photo-Optical
  Instrumentation Engineers (SPIE) Conference Series}}},\ \bibinfo {series}
  {Society of Photo-Optical Instrumentation Engineers (SPIE) Conference
  Series}, Vol.\ \bibinfo {volume} {7733}\ (\bibinfo {year} {2010})\BibitemShut
  {NoStop}%
\bibitem [{\citenamefont {{LSST Science Collaboration}}\ \emph
  {et~al.}(2009)\citenamefont {{LSST Science Collaboration}}, \citenamefont
  {{Abell}}, \citenamefont {{Allison}}, \citenamefont {{Anderson}},
  \citenamefont {{Andrew}}, \citenamefont {{Angel}}, \citenamefont {{Armus}},
  \citenamefont {{Arnett}}, \citenamefont {{Asztalos}}, \citenamefont
  {{Axelrod}},\ and\ \citenamefont {et~al.}}]{2009arXiv0912.0201L}%
  \BibitemOpen
  \bibfield  {author} {\bibinfo {author} {\bibnamefont {{LSST Science
  Collaboration}}}, \bibinfo {author} {\bibfnamefont {P.~A.}\ \bibnamefont
  {{Abell}}}, \bibinfo {author} {\bibfnamefont {J.}~\bibnamefont {{Allison}}},
  \bibinfo {author} {\bibfnamefont {S.~F.}\ \bibnamefont {{Anderson}}},
  \bibinfo {author} {\bibfnamefont {J.~R.}\ \bibnamefont {{Andrew}}}, \bibinfo
  {author} {\bibfnamefont {J.~R.~P.}\ \bibnamefont {{Angel}}}, \bibinfo
  {author} {\bibfnamefont {L.}~\bibnamefont {{Armus}}}, \bibinfo {author}
  {\bibfnamefont {D.}~\bibnamefont {{Arnett}}}, \bibinfo {author}
  {\bibfnamefont {S.~J.}\ \bibnamefont {{Asztalos}}}, \bibinfo {author}
  {\bibfnamefont {T.~S.}\ \bibnamefont {{Axelrod}}}, \ and\ \bibinfo {author}
  {\bibnamefont {et~al.}},\ }\href@noop {} {\bibfield  {journal} {\bibinfo
  {journal} {ArXiv e-prints}\ } (\bibinfo {year} {2009})},\ \Eprint
  {http://arxiv.org/abs/0912.0201} {arXiv:0912.0201 [astro-ph.IM]} \BibitemShut
  {NoStop}%
\bibitem [{\citenamefont {Schnittman}(2013)}]{Schnittman:2013qxa}%
  \BibitemOpen
  \bibfield  {author} {\bibinfo {author} {\bibfnamefont {J.~D.}\ \bibnamefont
  {Schnittman}},\ }\href {\doibase 10.1088/0264-9381/30/24/244007} {\bibfield
  {journal} {\bibinfo  {journal} {Class. Quantum Grav.}\ }\textbf {\bibinfo
  {volume} {30}},\ \bibinfo {pages} {244007} (\bibinfo {year} {2013})},\
  \Eprint {http://arxiv.org/abs/1307.3542} {arXiv:1307.3542 [gr-qc]}
  \BibitemShut {NoStop}%
\bibitem [{\citenamefont {Pretorius}(2005)}]{Pretorius:2005gq}%
  \BibitemOpen
  \bibfield  {author} {\bibinfo {author} {\bibfnamefont {F.}~\bibnamefont
  {Pretorius}},\ }\href {\doibase 10.1103/PhysRevLett.95.121101} {\bibfield
  {journal} {\bibinfo  {journal} {Phys. Rev. Lett.}\ }\textbf {\bibinfo
  {volume} {95}},\ \bibinfo {pages} {121101} (\bibinfo {year} {2005})},\
  \Eprint {http://arxiv.org/abs/gr-qc/0507014} {arXiv:gr-qc/0507014}
  \BibitemShut {NoStop}%
\bibitem [{\citenamefont {Campanelli}\ \emph {et~al.}(2006)\citenamefont
  {Campanelli}, \citenamefont {Lousto}, \citenamefont {Marronetti},\ and\
  \citenamefont {Zlochower}}]{Campanelli:2005dd}%
  \BibitemOpen
  \bibfield  {author} {\bibinfo {author} {\bibfnamefont {M.}~\bibnamefont
  {Campanelli}}, \bibinfo {author} {\bibfnamefont {C.~O.}\ \bibnamefont
  {Lousto}}, \bibinfo {author} {\bibfnamefont {P.}~\bibnamefont {Marronetti}},
  \ and\ \bibinfo {author} {\bibfnamefont {Y.}~\bibnamefont {Zlochower}},\
  }\href@noop {} {\bibfield  {journal} {\bibinfo  {journal} {\prl}\ }\textbf
  {\bibinfo {volume} {96}},\ \bibinfo {pages} {111101} (\bibinfo {year}
  {2006})},\ \Eprint {http://arxiv.org/abs/gr-qc/0511048} {gr-qc/0511048}
  \BibitemShut {NoStop}%
\bibitem [{\citenamefont {Baker}\ \emph {et~al.}(2006)\citenamefont {Baker},
  \citenamefont {Centrella}, \citenamefont {Choi}, \citenamefont {Koppitz},\
  and\ \citenamefont {van Meter}}]{Baker:2005vv}%
  \BibitemOpen
  \bibfield  {author} {\bibinfo {author} {\bibfnamefont {J.~G.}\ \bibnamefont
  {Baker}}, \bibinfo {author} {\bibfnamefont {J.}~\bibnamefont {Centrella}},
  \bibinfo {author} {\bibfnamefont {D.-I.}\ \bibnamefont {Choi}}, \bibinfo
  {author} {\bibfnamefont {M.}~\bibnamefont {Koppitz}}, \ and\ \bibinfo
  {author} {\bibfnamefont {J.}~\bibnamefont {van Meter}},\ }\href@noop {}
  {\bibfield  {journal} {\bibinfo  {journal} {\prl}\ }\textbf {\bibinfo
  {volume} {96}},\ \bibinfo {pages} {111102} (\bibinfo {year} {2006})},\
  \Eprint {http://arxiv.org/abs/gr-qc/0511103} {gr-qc/0511103} \BibitemShut
  {NoStop}%
\bibitem [{\citenamefont {Farris}\ \emph {et~al.}(2011)\citenamefont {Farris},
  \citenamefont {Liu},\ and\ \citenamefont {Shapiro}}]{Farris:2011vx}%
  \BibitemOpen
  \bibfield  {author} {\bibinfo {author} {\bibfnamefont {B.~D.}\ \bibnamefont
  {Farris}}, \bibinfo {author} {\bibfnamefont {Y.~T.}\ \bibnamefont {Liu}}, \
  and\ \bibinfo {author} {\bibfnamefont {S.~L.}\ \bibnamefont {Shapiro}},\
  }\href {\doibase 10.1103/PhysRevD.84.024024} {\bibfield  {journal} {\bibinfo
  {journal} {Phys. Rev.}\ }\textbf {\bibinfo {volume} {D84}},\ \bibinfo {pages}
  {024024} (\bibinfo {year} {2011})},\ \Eprint {http://arxiv.org/abs/1105.2821}
  {arXiv:1105.2821 [astro-ph.HE]} \BibitemShut {NoStop}%
\bibitem [{\citenamefont {{Bode}}\ \emph {et~al.}(2012)\citenamefont {{Bode}},
  \citenamefont {{Bogdanovi{\'c}}}, \citenamefont {{Haas}}, \citenamefont
  {{Healy}}, \citenamefont {{Laguna}},\ and\ \citenamefont
  {{Shoemaker}}}]{Bode12}%
  \BibitemOpen
  \bibfield  {author} {\bibinfo {author} {\bibfnamefont {T.}~\bibnamefont
  {{Bode}}}, \bibinfo {author} {\bibfnamefont {T.}~\bibnamefont
  {{Bogdanovi{\'c}}}}, \bibinfo {author} {\bibfnamefont {R.}~\bibnamefont
  {{Haas}}}, \bibinfo {author} {\bibfnamefont {J.}~\bibnamefont {{Healy}}},
  \bibinfo {author} {\bibfnamefont {P.}~\bibnamefont {{Laguna}}}, \ and\
  \bibinfo {author} {\bibfnamefont {D.}~\bibnamefont {{Shoemaker}}},\ }\href
  {\doibase 10.1088/0004-637X/744/1/45} {\bibfield  {journal} {\bibinfo
  {journal} {\apj}\ }\textbf {\bibinfo {volume} {744}},\ \bibinfo {pages} {45}
  (\bibinfo {year} {2012})},\ \Eprint {http://arxiv.org/abs/1101.4684}
  {arXiv:1101.4684 [gr-qc]} \BibitemShut {NoStop}%
\bibitem [{\citenamefont {Giacomazzo}\ \emph {et~al.}(2012)\citenamefont
  {Giacomazzo}, \citenamefont {Baker}, \citenamefont {Miller}, \citenamefont
  {Reynolds},\ and\ \citenamefont {van Meter}}]{Giacomazzo:2012iv}%
  \BibitemOpen
  \bibfield  {author} {\bibinfo {author} {\bibfnamefont {B.}~\bibnamefont
  {Giacomazzo}}, \bibinfo {author} {\bibfnamefont {J.~G.}\ \bibnamefont
  {Baker}}, \bibinfo {author} {\bibfnamefont {M.~C.}\ \bibnamefont {Miller}},
  \bibinfo {author} {\bibfnamefont {C.~S.}\ \bibnamefont {Reynolds}}, \ and\
  \bibinfo {author} {\bibfnamefont {J.~R.}\ \bibnamefont {van Meter}},\
  }\href@noop {} {\bibfield  {journal} {\bibinfo  {journal} {Astrophys. J.}\
  }\textbf {\bibinfo {volume} {752}},\ \bibinfo {pages} {L15} (\bibinfo {year}
  {2012})},\ \Eprint {http://arxiv.org/abs/1203.6108} {arXiv:1203.6108
  [astro-ph.HE]} \BibitemShut {NoStop}%
\bibitem [{\citenamefont {Blanchet}(2006)}]{Blanchet:2006zz}%
  \BibitemOpen
  \bibfield  {author} {\bibinfo {author} {\bibfnamefont {L.}~\bibnamefont
  {Blanchet}},\ }\href@noop {} {\bibfield  {journal} {\bibinfo  {journal}
  {Living Rev. Relativity}\ }\textbf {\bibinfo {volume} {9}},\ \bibinfo {pages}
  {4} (\bibinfo {year} {2006})}\BibitemShut {NoStop}%
\bibitem [{\citenamefont {Noble}\ \emph {et~al.}(2012)\citenamefont {Noble},
  \citenamefont {Mundim}, \citenamefont {Nakano}, \citenamefont {Krolik},
  \citenamefont {Campanelli} \emph {et~al.}}]{Noble:2012xz}%
  \BibitemOpen
  \bibfield  {author} {\bibinfo {author} {\bibfnamefont {S.~C.}\ \bibnamefont
  {Noble}}, \bibinfo {author} {\bibfnamefont {B.~C.}\ \bibnamefont {Mundim}},
  \bibinfo {author} {\bibfnamefont {H.}~\bibnamefont {Nakano}}, \bibinfo
  {author} {\bibfnamefont {J.~H.}\ \bibnamefont {Krolik}}, \bibinfo {author}
  {\bibfnamefont {M.}~\bibnamefont {Campanelli}},  \emph {et~al.},\ }\href
  {\doibase 10.1088/0004-637X/755/1/51} {\bibfield  {journal} {\bibinfo
  {journal} {Astrophys. J.}\ }\textbf {\bibinfo {volume} {755}},\ \bibinfo
  {pages} {51} (\bibinfo {year} {2012})},\ \Eprint
  {http://arxiv.org/abs/1204.1073} {arXiv:1204.1073 [astro-ph.HE]} \BibitemShut
  {NoStop}%
\bibitem [{\citenamefont {{Noble}}\ \emph {et~al.}(2009)\citenamefont
  {{Noble}}, \citenamefont {{Krolik}},\ and\ \citenamefont
  {{Hawley}}}]{Noble09}%
  \BibitemOpen
  \bibfield  {author} {\bibinfo {author} {\bibfnamefont {S.~C.}\ \bibnamefont
  {{Noble}}}, \bibinfo {author} {\bibfnamefont {J.~H.}\ \bibnamefont
  {{Krolik}}}, \ and\ \bibinfo {author} {\bibfnamefont {J.~F.}\ \bibnamefont
  {{Hawley}}},\ }\href {\doibase 10.1088/0004-637X/692/1/411} {\bibfield
  {journal} {\bibinfo  {journal} {\apj}\ }\textbf {\bibinfo {volume} {692}},\
  \bibinfo {pages} {411} (\bibinfo {year} {2009})},\ \Eprint
  {http://arxiv.org/abs/0808.3140} {arXiv:0808.3140} \BibitemShut {NoStop}%
\bibitem [{\citenamefont {Shi}\ \emph {et~al.}(2012)\citenamefont {Shi},
  \citenamefont {Krolik}, \citenamefont {Lubow},\ and\ \citenamefont
  {Hawley}}]{Shi:2011us}%
  \BibitemOpen
  \bibfield  {author} {\bibinfo {author} {\bibfnamefont {J.-M.}\ \bibnamefont
  {Shi}}, \bibinfo {author} {\bibfnamefont {J.~H.}\ \bibnamefont {Krolik}},
  \bibinfo {author} {\bibfnamefont {S.~H.}\ \bibnamefont {Lubow}}, \ and\
  \bibinfo {author} {\bibfnamefont {J.~F.}\ \bibnamefont {Hawley}},\ }\href
  {\doibase 10.1088/0004-637X/749/2/118} {\bibfield  {journal} {\bibinfo
  {journal} {Astrophys. J.}\ }\textbf {\bibinfo {volume} {749}},\ \bibinfo
  {pages} {118} (\bibinfo {year} {2012})},\ \Eprint
  {http://arxiv.org/abs/1110.4866} {arXiv:1110.4866 [astro-ph.HE]} \BibitemShut
  {NoStop}%
\bibitem [{\citenamefont {Yunes}\ \emph {et~al.}(2006)\citenamefont {Yunes},
  \citenamefont {Tichy}, \citenamefont {Owen},\ and\ \citenamefont
  {Bruegmann}}]{Yunes:2005nn}%
  \BibitemOpen
  \bibfield  {author} {\bibinfo {author} {\bibfnamefont {N.}~\bibnamefont
  {Yunes}}, \bibinfo {author} {\bibfnamefont {W.}~\bibnamefont {Tichy}},
  \bibinfo {author} {\bibfnamefont {B.~J.}\ \bibnamefont {Owen}}, \ and\
  \bibinfo {author} {\bibfnamefont {B.}~\bibnamefont {Bruegmann}},\ }\href
  {\doibase 10.1103/PhysRevD.74.104011} {\bibfield  {journal} {\bibinfo
  {journal} {Phys. Rev.}\ }\textbf {\bibinfo {volume} {D74}},\ \bibinfo {pages}
  {104011} (\bibinfo {year} {2006})},\ \Eprint
  {http://arxiv.org/abs/gr-qc/0503011} {arXiv:gr-qc/0503011} \BibitemShut
  {NoStop}%
\bibitem [{\citenamefont {Thorne}(1980)}]{Thorne:1980ru}%
  \BibitemOpen
  \bibfield  {author} {\bibinfo {author} {\bibfnamefont {K.~S.}\ \bibnamefont
  {Thorne}},\ }\href {\doibase 10.1103/RevModPhys.52.299} {\bibfield  {journal}
  {\bibinfo  {journal} {Rev. Mod. Phys.}\ }\textbf {\bibinfo {volume} {52}},\
  \bibinfo {pages} {299} (\bibinfo {year} {1980})}\BibitemShut {NoStop}%
\bibitem [{\citenamefont {Thorne}\ and\ \citenamefont
  {Hartle}(1985)}]{Thorne:1984mz}%
  \BibitemOpen
  \bibfield  {author} {\bibinfo {author} {\bibfnamefont {K.~S.}\ \bibnamefont
  {Thorne}}\ and\ \bibinfo {author} {\bibfnamefont {J.~B.}\ \bibnamefont
  {Hartle}},\ }\href {\doibase 10.1103/PhysRevD.31.1815} {\bibfield  {journal}
  {\bibinfo  {journal} {Phys. Rev.}\ }\textbf {\bibinfo {volume} {D31}},\
  \bibinfo {pages} {1815} (\bibinfo {year} {1985})}\BibitemShut {NoStop}%
\bibitem [{\citenamefont {Detweiler}(2005)}]{Detweiler:2005kq}%
  \BibitemOpen
  \bibfield  {author} {\bibinfo {author} {\bibfnamefont {S.~L.}\ \bibnamefont
  {Detweiler}},\ }\href {\doibase 10.1088/0264-9381/22/15/006} {\bibfield
  {journal} {\bibinfo  {journal} {Class. Quantum Grav.}\ }\textbf {\bibinfo
  {volume} {22}},\ \bibinfo {pages} {S681} (\bibinfo {year} {2005})},\ \Eprint
  {http://arxiv.org/abs/gr-qc/0501004} {arXiv:gr-qc/0501004} \BibitemShut
  {NoStop}%
\bibitem [{\citenamefont {Yunes}\ and\ \citenamefont
  {Gonzalez}(2006)}]{Yunes:2005ve}%
  \BibitemOpen
  \bibfield  {author} {\bibinfo {author} {\bibfnamefont {N.}~\bibnamefont
  {Yunes}}\ and\ \bibinfo {author} {\bibfnamefont {J.~A.}\ \bibnamefont
  {Gonzalez}},\ }\href {\doibase 10.1103/PhysRevD.73.024010} {\bibfield
  {journal} {\bibinfo  {journal} {Phys. Rev.}\ }\textbf {\bibinfo {volume}
  {D73}},\ \bibinfo {pages} {024010} (\bibinfo {year} {2006})},\ \Eprint
  {http://arxiv.org/abs/gr-qc/0510076} {arXiv:gr-qc/0510076} \BibitemShut
  {NoStop}%
\bibitem [{\citenamefont {Chatziioannou}\ \emph {et~al.}(2013)\citenamefont
  {Chatziioannou}, \citenamefont {Poisson},\ and\ \citenamefont
  {Yunes}}]{Chatziioannou:2012gq}%
  \BibitemOpen
  \bibfield  {author} {\bibinfo {author} {\bibfnamefont {K.}~\bibnamefont
  {Chatziioannou}}, \bibinfo {author} {\bibfnamefont {E.}~\bibnamefont
  {Poisson}}, \ and\ \bibinfo {author} {\bibfnamefont {N.}~\bibnamefont
  {Yunes}},\ }\href {\doibase 10.1103/PhysRevD.87.044022} {\bibfield  {journal}
  {\bibinfo  {journal} {Phys. Rev.}\ }\textbf {\bibinfo {volume} {D87}},\
  \bibinfo {pages} {044022} (\bibinfo {year} {2013})},\ \Eprint
  {http://arxiv.org/abs/1211.1686} {arXiv:1211.1686 [gr-qc]} \BibitemShut
  {NoStop}%
\bibitem [{\citenamefont {Blanchet}\ \emph {et~al.}(1998)\citenamefont
  {Blanchet}, \citenamefont {Faye},\ and\ \citenamefont
  {Ponsot}}]{Blanchet:1998vx}%
  \BibitemOpen
  \bibfield  {author} {\bibinfo {author} {\bibfnamefont {L.}~\bibnamefont
  {Blanchet}}, \bibinfo {author} {\bibfnamefont {G.}~\bibnamefont {Faye}}, \
  and\ \bibinfo {author} {\bibfnamefont {B.}~\bibnamefont {Ponsot}},\ }\href
  {\doibase 10.1103/PhysRevD.58.124002} {\bibfield  {journal} {\bibinfo
  {journal} {Phys. Rev.}\ }\textbf {\bibinfo {volume} {D58}},\ \bibinfo {pages}
  {124002} (\bibinfo {year} {1998})},\ \Eprint
  {http://arxiv.org/abs/gr-qc/9804079} {arXiv:gr-qc/9804079} \BibitemShut
  {NoStop}%
\bibitem [{\citenamefont {Will}\ and\ \citenamefont
  {Wiseman}(1996)}]{Will:1996zj}%
  \BibitemOpen
  \bibfield  {author} {\bibinfo {author} {\bibfnamefont {C.~M.}\ \bibnamefont
  {Will}}\ and\ \bibinfo {author} {\bibfnamefont {A.~G.}\ \bibnamefont
  {Wiseman}},\ }\href {\doibase 10.1103/PhysRevD.54.4813} {\bibfield  {journal}
  {\bibinfo  {journal} {Phys. Rev.}\ }\textbf {\bibinfo {volume} {D54}},\
  \bibinfo {pages} {4813} (\bibinfo {year} {1996})},\ \Eprint
  {http://arxiv.org/abs/gr-qc/9608012} {arXiv:gr-qc/9608012} \BibitemShut
  {NoStop}%
\bibitem [{\citenamefont {Alvi}(2000)}]{Alvi:1999cw}%
  \BibitemOpen
  \bibfield  {author} {\bibinfo {author} {\bibfnamefont {K.}~\bibnamefont
  {Alvi}},\ }\href {\doibase 10.1103/PhysRevD.61.124013} {\bibfield  {journal}
  {\bibinfo  {journal} {Phys. Rev.}\ }\textbf {\bibinfo {volume} {D61}},\
  \bibinfo {pages} {124013} (\bibinfo {year} {2000})},\ \Eprint
  {http://arxiv.org/abs/gr-qc/9912113} {arXiv:gr-qc/9912113} \BibitemShut
  {NoStop}%
\bibitem [{\citenamefont {Pati}\ and\ \citenamefont
  {Will}(2000)}]{Pati:2000vt}%
  \BibitemOpen
  \bibfield  {author} {\bibinfo {author} {\bibfnamefont {M.~E.}\ \bibnamefont
  {Pati}}\ and\ \bibinfo {author} {\bibfnamefont {C.~M.}\ \bibnamefont
  {Will}},\ }\href {\doibase 10.1103/PhysRevD.62.124015} {\bibfield  {journal}
  {\bibinfo  {journal} {Phys. Rev.}\ }\textbf {\bibinfo {volume} {D62}},\
  \bibinfo {pages} {124015} (\bibinfo {year} {2000})},\ \Eprint
  {http://arxiv.org/abs/gr-qc/0007087} {arXiv:gr-qc/0007087} \BibitemShut
  {NoStop}%
\bibitem [{\citenamefont {Pati}\ and\ \citenamefont
  {Will}(2002)}]{Pati:2002ux}%
  \BibitemOpen
  \bibfield  {author} {\bibinfo {author} {\bibfnamefont {M.~E.}\ \bibnamefont
  {Pati}}\ and\ \bibinfo {author} {\bibfnamefont {C.~M.}\ \bibnamefont
  {Will}},\ }\href {\doibase 10.1103/PhysRevD.65.104008} {\bibfield  {journal}
  {\bibinfo  {journal} {Phys. Rev.}\ }\textbf {\bibinfo {volume} {D65}},\
  \bibinfo {pages} {104008} (\bibinfo {year} {2002})},\ \Eprint
  {http://arxiv.org/abs/gr-qc/0201001} {arXiv:gr-qc/0201001} \BibitemShut
  {NoStop}%
\bibitem [{\citenamefont {Alvi}(2003)}]{Alvi:2003pn}%
  \BibitemOpen
  \bibfield  {author} {\bibinfo {author} {\bibfnamefont {K.}~\bibnamefont
  {Alvi}},\ }\href {\doibase 10.1103/PhysRevD.67.104006} {\bibfield  {journal}
  {\bibinfo  {journal} {Phys. Rev.}\ }\textbf {\bibinfo {volume} {D67}},\
  \bibinfo {pages} {104006} (\bibinfo {year} {2003})},\ \Eprint
  {http://arxiv.org/abs/gr-qc/0302061} {arXiv:gr-qc/0302061} \BibitemShut
  {NoStop}%
\bibitem [{\citenamefont {Yunes}\ and\ \citenamefont
  {Tichy}(2006)}]{Yunes:2006iw}%
  \BibitemOpen
  \bibfield  {author} {\bibinfo {author} {\bibfnamefont {N.}~\bibnamefont
  {Yunes}}\ and\ \bibinfo {author} {\bibfnamefont {W.}~\bibnamefont {Tichy}},\
  }\href {\doibase 10.1103/PhysRevD.74.064013} {\bibfield  {journal} {\bibinfo
  {journal} {Phys. Rev.}\ }\textbf {\bibinfo {volume} {D74}},\ \bibinfo {pages}
  {064013} (\bibinfo {year} {2006})},\ \Eprint
  {http://arxiv.org/abs/gr-qc/0601046} {arXiv:gr-qc/0601046} \BibitemShut
  {NoStop}%
\bibitem [{\citenamefont {Yunes}(2007)}]{Yunes:2006mx}%
  \BibitemOpen
  \bibfield  {author} {\bibinfo {author} {\bibfnamefont {N.}~\bibnamefont
  {Yunes}},\ }\href {\doibase 10.1088/0264-9381/24/17/004} {\bibfield
  {journal} {\bibinfo  {journal} {Class. Quantum Grav.}\ }\textbf {\bibinfo
  {volume} {24}},\ \bibinfo {pages} {4313} (\bibinfo {year} {2007})},\ \Eprint
  {http://arxiv.org/abs/gr-qc/0611128} {arXiv:gr-qc/0611128} \BibitemShut
  {NoStop}%
\bibitem [{\citenamefont {Johnson-McDaniel}\ \emph {et~al.}(2009)\citenamefont
  {Johnson-McDaniel}, \citenamefont {Yunes}, \citenamefont {Tichy},\ and\
  \citenamefont {Owen}}]{JohnsonMcDaniel:2009dq}%
  \BibitemOpen
  \bibfield  {author} {\bibinfo {author} {\bibfnamefont {N.~K.}\ \bibnamefont
  {Johnson-McDaniel}}, \bibinfo {author} {\bibfnamefont {N.}~\bibnamefont
  {Yunes}}, \bibinfo {author} {\bibfnamefont {W.}~\bibnamefont {Tichy}}, \ and\
  \bibinfo {author} {\bibfnamefont {B.~J.}\ \bibnamefont {Owen}},\ }\href
  {\doibase 10.1103/PhysRevD.80.124039} {\bibfield  {journal} {\bibinfo
  {journal} {Phys. Rev.}\ }\textbf {\bibinfo {volume} {D80}},\ \bibinfo {pages}
  {124039} (\bibinfo {year} {2009})},\ \Eprint {http://arxiv.org/abs/0907.0891}
  {arXiv:0907.0891 [gr-qc]} \BibitemShut {NoStop}%
\bibitem [{\citenamefont {Misner}\ \emph {et~al.}(1973)\citenamefont {Misner},
  \citenamefont {Thorne},\ and\ \citenamefont {Wheeler}}]{MTW}%
  \BibitemOpen
  \bibfield  {author} {\bibinfo {author} {\bibfnamefont {C.~W.}\ \bibnamefont
  {Misner}}, \bibinfo {author} {\bibfnamefont {K.}~\bibnamefont {Thorne}}, \
  and\ \bibinfo {author} {\bibfnamefont {J.~A.}\ \bibnamefont {Wheeler}},\
  }\href@noop {} {\emph {\bibinfo {title} {Gravitation}}}\ (\bibinfo
  {publisher} {W. H. Freeman \& Co.},\ \bibinfo {address} {San Francisco},\
  \bibinfo {year} {1973})\BibitemShut {NoStop}%
\bibitem [{\citenamefont {Gallouin}\ \emph {et~al.}(2012)\citenamefont
  {Gallouin}, \citenamefont {Nakano}, \citenamefont {Yunes},\ and\
  \citenamefont {Campanelli}}]{Gallouin:2012kb}%
  \BibitemOpen
  \bibfield  {author} {\bibinfo {author} {\bibfnamefont {L.}~\bibnamefont
  {Gallouin}}, \bibinfo {author} {\bibfnamefont {H.}~\bibnamefont {Nakano}},
  \bibinfo {author} {\bibfnamefont {N.}~\bibnamefont {Yunes}}, \ and\ \bibinfo
  {author} {\bibfnamefont {M.}~\bibnamefont {Campanelli}},\ }\href {\doibase
  10.1088/0264-9381/29/23/235013} {\bibfield  {journal} {\bibinfo  {journal}
  {Class. Quantum Grav.}\ }\textbf {\bibinfo {volume} {29}},\ \bibinfo {pages}
  {235013} (\bibinfo {year} {2012})},\ \Eprint {http://arxiv.org/abs/1208.6489}
  {arXiv:1208.6489 [gr-qc]} \BibitemShut {NoStop}%
\bibitem [{\citenamefont {Poisson}(2005)}]{Poisson:2005pi}%
  \BibitemOpen
  \bibfield  {author} {\bibinfo {author} {\bibfnamefont {E.}~\bibnamefont
  {Poisson}},\ }\href {\doibase 10.1103/PhysRevLett.94.161103} {\bibfield
  {journal} {\bibinfo  {journal} {Phys. Rev. Lett.}\ }\textbf {\bibinfo
  {volume} {94}},\ \bibinfo {pages} {161103} (\bibinfo {year} {2005})},\
  \Eprint {http://arxiv.org/abs/gr-qc/0501032} {arXiv:gr-qc/0501032}
  \BibitemShut {NoStop}%
\bibitem [{\citenamefont {Taylor}\ and\ \citenamefont
  {Poisson}(2008)}]{Taylor:2008xy}%
  \BibitemOpen
  \bibfield  {author} {\bibinfo {author} {\bibfnamefont {S.}~\bibnamefont
  {Taylor}}\ and\ \bibinfo {author} {\bibfnamefont {E.}~\bibnamefont
  {Poisson}},\ }\href {\doibase 10.1103/PhysRevD.78.084016} {\bibfield
  {journal} {\bibinfo  {journal} {Phys. Rev.}\ }\textbf {\bibinfo {volume}
  {D78}},\ \bibinfo {pages} {084016} (\bibinfo {year} {2008})},\ \Eprint
  {http://arxiv.org/abs/0806.3052} {arXiv:0806.3052 [gr-qc]} \BibitemShut
  {NoStop}%
\bibitem [{\citenamefont {Comeau}\ and\ \citenamefont
  {Poisson}(2009)}]{Comeau:2009bz}%
  \BibitemOpen
  \bibfield  {author} {\bibinfo {author} {\bibfnamefont {S.}~\bibnamefont
  {Comeau}}\ and\ \bibinfo {author} {\bibfnamefont {E.}~\bibnamefont
  {Poisson}},\ }\href {\doibase 10.1103/PhysRevD.80.087501} {\bibfield
  {journal} {\bibinfo  {journal} {Phys. Rev.}\ }\textbf {\bibinfo {volume}
  {D80}},\ \bibinfo {pages} {087501} (\bibinfo {year} {2009})},\ \Eprint
  {http://arxiv.org/abs/0908.4518} {arXiv:0908.4518 [gr-qc]} \BibitemShut
  {NoStop}%
\bibitem [{\citenamefont {Poisson}\ and\ \citenamefont
  {Vlasov}(2010)}]{Poisson:2009qj}%
  \BibitemOpen
  \bibfield  {author} {\bibinfo {author} {\bibfnamefont {E.}~\bibnamefont
  {Poisson}}\ and\ \bibinfo {author} {\bibfnamefont {I.}~\bibnamefont
  {Vlasov}},\ }\href {\doibase 10.1103/PhysRevD.81.024029} {\bibfield
  {journal} {\bibinfo  {journal} {Phys. Rev.}\ }\textbf {\bibinfo {volume}
  {D81}},\ \bibinfo {pages} {024029} (\bibinfo {year} {2010})},\ \Eprint
  {http://arxiv.org/abs/0910.4311} {arXiv:0910.4311 [gr-qc]} \BibitemShut
  {NoStop}%
\bibitem [{\citenamefont {Cook}\ and\ \citenamefont
  {Scheel}(1997)}]{Cook:1997qc}%
  \BibitemOpen
  \bibfield  {author} {\bibinfo {author} {\bibfnamefont {G.~B.}\ \bibnamefont
  {Cook}}\ and\ \bibinfo {author} {\bibfnamefont {M.~A.}\ \bibnamefont
  {Scheel}},\ }\href {\doibase 10.1103/PhysRevD.56.4775} {\bibfield  {journal}
  {\bibinfo  {journal} {Phys. Rev.}\ }\textbf {\bibinfo {volume} {D56}},\
  \bibinfo {pages} {4775} (\bibinfo {year} {1997})}\BibitemShut {NoStop}%
\bibitem [{\citenamefont {Blanchet}\ \emph {et~al.}(1995)\citenamefont
  {Blanchet}, \citenamefont {Damour},\ and\ \citenamefont
  {Iyer}}]{Blanchet:1995fg}%
  \BibitemOpen
  \bibfield  {author} {\bibinfo {author} {\bibfnamefont {L.}~\bibnamefont
  {Blanchet}}, \bibinfo {author} {\bibfnamefont {T.}~\bibnamefont {Damour}}, \
  and\ \bibinfo {author} {\bibfnamefont {B.~R.}\ \bibnamefont {Iyer}},\ }\href
  {\doibase 10.1103/PhysRevD.51.5360, 10.1103/PhysRevD.54.1860} {\bibfield
  {journal} {\bibinfo  {journal} {Phys. Rev.}\ }\textbf {\bibinfo {volume}
  {D51}},\ \bibinfo {pages} {5360} (\bibinfo {year} {1995})},\ \Eprint
  {http://arxiv.org/abs/gr-qc/9501029} {arXiv:gr-qc/9501029} \BibitemShut
  {NoStop}%
\bibitem [{\citenamefont {Blanchet}(1995)}]{Blanchet:1995fr}%
  \BibitemOpen
  \bibfield  {author} {\bibinfo {author} {\bibfnamefont {L.}~\bibnamefont
  {Blanchet}},\ }\href {\doibase 10.1103/PhysRevD.51.2559} {\bibfield
  {journal} {\bibinfo  {journal} {Phys. Rev.}\ }\textbf {\bibinfo {volume}
  {D51}},\ \bibinfo {pages} {2559} (\bibinfo {year} {1995})},\ \Eprint
  {http://arxiv.org/abs/gr-qc/9501030} {arXiv:gr-qc/9501030} \BibitemShut
  {NoStop}%
\bibitem [{\citenamefont {Blanchet}(1996)}]{Blanchet:1996wx}%
  \BibitemOpen
  \bibfield  {author} {\bibinfo {author} {\bibfnamefont {L.}~\bibnamefont
  {Blanchet}},\ }\href {\doibase 10.1103/PhysRevD.71.129904,
  10.1103/PhysRevD.54.1417} {\bibfield  {journal} {\bibinfo  {journal} {Phys.
  Rev.}\ }\textbf {\bibinfo {volume} {D54}},\ \bibinfo {pages} {1417} (\bibinfo
  {year} {1996})},\ \Eprint {http://arxiv.org/abs/gr-qc/9603048}
  {arXiv:gr-qc/9603048} \BibitemShut {NoStop}%
\bibitem [{\citenamefont {Bender}\ and\ \citenamefont {Orszag}(1999)}]{Bender}%
  \BibitemOpen
  \bibfield  {author} {\bibinfo {author} {\bibfnamefont {C.~M.}\ \bibnamefont
  {Bender}}\ and\ \bibinfo {author} {\bibfnamefont {S.~A.}\ \bibnamefont
  {Orszag}},\ }\href@noop {} {\emph {\bibinfo {title} {Advanced mathematical
  methods for scientists and engineers 1, Asymptotic methods and perturbation
  theory}}}\ (\bibinfo  {publisher} {Springer},\ \bibinfo {address} {New
  York},\ \bibinfo {year} {1999})\BibitemShut {NoStop}%
\bibitem [{\citenamefont {Reifenberger}\ and\ \citenamefont
  {Tichy}(2012)}]{Reifenberger:2012yg}%
  \BibitemOpen
  \bibfield  {author} {\bibinfo {author} {\bibfnamefont {G.}~\bibnamefont
  {Reifenberger}}\ and\ \bibinfo {author} {\bibfnamefont {W.}~\bibnamefont
  {Tichy}},\ }\href {\doibase 10.1103/PhysRevD.86.064003} {\bibfield  {journal}
  {\bibinfo  {journal} {Phys. Rev.}\ }\textbf {\bibinfo {volume} {D86}},\
  \bibinfo {pages} {064003} (\bibinfo {year} {2012})},\ \Eprint
  {http://arxiv.org/abs/1205.5502} {arXiv:1205.5502 [gr-qc]} \BibitemShut
  {NoStop}%
\bibitem [{\citenamefont {Chu}(2012)}]{Chu:thesis}%
  \BibitemOpen
  \bibfield  {author} {\bibinfo {author} {\bibfnamefont {T.}~\bibnamefont
  {Chu}},\ }\emph {\bibinfo {title} {Numerical simulations of black-hole
  spacetimes}},\ \href {http://thesis.library.caltech.edu/6556/} {Ph.D.
  thesis},\ \bibinfo  {school} {California Institute of Technology}, \bibinfo
  {address} {Pasadena, CA 91125, USA} (\bibinfo {year} {2012})\BibitemShut
  {NoStop}%
\bibitem [{\citenamefont {Poisson}(2004{\natexlab{a}})}]{Poisson:2003wz}%
  \BibitemOpen
  \bibfield  {author} {\bibinfo {author} {\bibfnamefont {E.}~\bibnamefont
  {Poisson}},\ }\href {\doibase 10.1103/PhysRevD.69.084007} {\bibfield
  {journal} {\bibinfo  {journal} {Phys. Rev.}\ }\textbf {\bibinfo {volume}
  {D69}},\ \bibinfo {pages} {084007} (\bibinfo {year} {2004}{\natexlab{a}})},\
  \Eprint {http://arxiv.org/abs/gr-qc/0311026} {arXiv:gr-qc/0311026}
  \BibitemShut {NoStop}%
\bibitem [{\citenamefont {Poisson}(2004{\natexlab{b}})}]{Poisson:2004cw}%
  \BibitemOpen
  \bibfield  {author} {\bibinfo {author} {\bibfnamefont {E.}~\bibnamefont
  {Poisson}},\ }\href {\doibase 10.1103/PhysRevD.70.084044} {\bibfield
  {journal} {\bibinfo  {journal} {Phys. Rev.}\ }\textbf {\bibinfo {volume}
  {D70}},\ \bibinfo {pages} {084044} (\bibinfo {year} {2004}{\natexlab{b}})},\
  \Eprint {http://arxiv.org/abs/gr-qc/0407050} {arXiv:gr-qc/0407050}
  \BibitemShut {NoStop}%
\bibitem [{\citenamefont {Poisson}\ \emph {et~al.}(2011)\citenamefont
  {Poisson}, \citenamefont {Pound},\ and\ \citenamefont
  {Vega}}]{Poisson:2011nh}%
  \BibitemOpen
  \bibfield  {author} {\bibinfo {author} {\bibfnamefont {E.}~\bibnamefont
  {Poisson}}, \bibinfo {author} {\bibfnamefont {A.}~\bibnamefont {Pound}}, \
  and\ \bibinfo {author} {\bibfnamefont {I.}~\bibnamefont {Vega}},\ }\href@noop
  {} {\bibfield  {journal} {\bibinfo  {journal} {Living Rev. Relativity}\
  }\textbf {\bibinfo {volume} {14}},\ \bibinfo {pages} {7} (\bibinfo {year}
  {2011})},\ \Eprint {http://arxiv.org/abs/1102.0529} {arXiv:1102.0529 [gr-qc]}
  \BibitemShut {NoStop}%
\bibitem [{\citenamefont {Arnowitt}\ \emph {et~al.}(2008)\citenamefont
  {Arnowitt}, \citenamefont {Deser},\ and\ \citenamefont
  {Misner}}]{Arnowitt:1962hi}%
  \BibitemOpen
  \bibfield  {author} {\bibinfo {author} {\bibfnamefont {R.~L.}\ \bibnamefont
  {Arnowitt}}, \bibinfo {author} {\bibfnamefont {S.}~\bibnamefont {Deser}}, \
  and\ \bibinfo {author} {\bibfnamefont {C.~W.}\ \bibnamefont {Misner}},\
  }\href {\doibase 10.1007/s10714-008-0661-1} {\bibfield  {journal} {\bibinfo
  {journal} {Gen. Rel. Grav.}\ }\textbf {\bibinfo {volume} {40}},\ \bibinfo
  {pages} {1997} (\bibinfo {year} {2008})},\ \Eprint
  {http://arxiv.org/abs/gr-qc/0405109} {arXiv:gr-qc/0405109} \BibitemShut
  {NoStop}%
\bibitem [{\citenamefont {{Nakamura}}\ \emph {et~al.}(1987)\citenamefont
  {{Nakamura}}, \citenamefont {{Oohara}},\ and\ \citenamefont
  {{Kojima}}}]{1987PThPS..90....1N}%
  \BibitemOpen
  \bibfield  {author} {\bibinfo {author} {\bibfnamefont {T.}~\bibnamefont
  {{Nakamura}}}, \bibinfo {author} {\bibfnamefont {K.}~\bibnamefont
  {{Oohara}}}, \ and\ \bibinfo {author} {\bibfnamefont {Y.}~\bibnamefont
  {{Kojima}}},\ }\href {\doibase 10.1143/PTPS.90.1} {\bibfield  {journal}
  {\bibinfo  {journal} {Prog. Theor. Phys. Suppl.}\ }\textbf {\bibinfo {volume}
  {90}},\ \bibinfo {pages} {1} (\bibinfo {year} {1987})}\BibitemShut {NoStop}%
\bibitem [{\citenamefont {{Noble}}\ \emph {et~al.}(2010)\citenamefont
  {{Noble}}, \citenamefont {{Krolik}},\ and\ \citenamefont
  {{Hawley}}}]{Noble10}%
  \BibitemOpen
  \bibfield  {author} {\bibinfo {author} {\bibfnamefont {S.~C.}\ \bibnamefont
  {{Noble}}}, \bibinfo {author} {\bibfnamefont {J.~H.}\ \bibnamefont
  {{Krolik}}}, \ and\ \bibinfo {author} {\bibfnamefont {J.~F.}\ \bibnamefont
  {{Hawley}}},\ }\href {\doibase 10.1088/0004-637X/711/2/959} {\bibfield
  {journal} {\bibinfo  {journal} {\apj}\ }\textbf {\bibinfo {volume} {711}},\
  \bibinfo {pages} {959} (\bibinfo {year} {2010})},\ \Eprint
  {http://arxiv.org/abs/1001.4809} {arXiv:1001.4809 [astro-ph.HE]} \BibitemShut
  {NoStop}%
\bibitem [{\citenamefont {Noble}\ \emph {et~al.}(2011)\citenamefont {Noble},
  \citenamefont {Krolik}, \citenamefont {Schnittman},\ and\ \citenamefont
  {Hawley}}]{Noble:2011wa}%
  \BibitemOpen
  \bibfield  {author} {\bibinfo {author} {\bibfnamefont {S.~C.}\ \bibnamefont
  {Noble}}, \bibinfo {author} {\bibfnamefont {J.~H.}\ \bibnamefont {Krolik}},
  \bibinfo {author} {\bibfnamefont {J.~D.}\ \bibnamefont {Schnittman}}, \ and\
  \bibinfo {author} {\bibfnamefont {J.~F.}\ \bibnamefont {Hawley}},\ }\href
  {\doibase 10.1088/0004-637X/743/2/115} {\bibfield  {journal} {\bibinfo
  {journal} {Astrophys. J.}\ }\textbf {\bibinfo {volume} {743}},\ \bibinfo
  {pages} {115} (\bibinfo {year} {2011})},\ \Eprint
  {http://arxiv.org/abs/1105.2825} {arXiv:1105.2825 [astro-ph.HE]} \BibitemShut
  {NoStop}%
\bibitem [{\citenamefont {Faye}\ \emph {et~al.}(2006)\citenamefont {Faye},
  \citenamefont {Blanchet},\ and\ \citenamefont {Buonanno}}]{Faye:2006gx}%
  \BibitemOpen
  \bibfield  {author} {\bibinfo {author} {\bibfnamefont {G.}~\bibnamefont
  {Faye}}, \bibinfo {author} {\bibfnamefont {L.}~\bibnamefont {Blanchet}}, \
  and\ \bibinfo {author} {\bibfnamefont {A.}~\bibnamefont {Buonanno}},\ }\href
  {\doibase 10.1103/PhysRevD.74.104033} {\bibfield  {journal} {\bibinfo
  {journal} {Phys. Rev.}\ }\textbf {\bibinfo {volume} {D74}},\ \bibinfo {pages}
  {104033} (\bibinfo {year} {2006})},\ \Eprint
  {http://arxiv.org/abs/gr-qc/0605139} {arXiv:gr-qc/0605139} \BibitemShut
  {NoStop}%
\bibitem [{\citenamefont {Damour}\ and\ \citenamefont
  {Deruelle}(1985)}]{Damour:1985}%
  \BibitemOpen
  \bibfield  {author} {\bibinfo {author} {\bibfnamefont {T.}~\bibnamefont
  {Damour}}\ and\ \bibinfo {author} {\bibfnamefont {N.}~\bibnamefont
  {Deruelle}},\ }\href@noop {} {\bibfield  {journal} {\bibinfo  {journal} {Ann.
  Inst. Henri Poincar{\'e} Phys. Th{\'e}or.,}\ }\textbf {\bibinfo {volume}
  {43}},\ \bibinfo {pages} {107} (\bibinfo {year} {1985})}\BibitemShut
  {NoStop}%
\bibitem [{\citenamefont {Gopakumar}\ and\ \citenamefont
  {Iyer}(2002)}]{Gopakumar:2001dy}%
  \BibitemOpen
  \bibfield  {author} {\bibinfo {author} {\bibfnamefont {A.}~\bibnamefont
  {Gopakumar}}\ and\ \bibinfo {author} {\bibfnamefont {B.~R.}\ \bibnamefont
  {Iyer}},\ }\href {\doibase 10.1103/PhysRevD.65.084011} {\bibfield  {journal}
  {\bibinfo  {journal} {Phys. Rev.}\ }\textbf {\bibinfo {volume} {D65}},\
  \bibinfo {pages} {084011} (\bibinfo {year} {2002})},\ \Eprint
  {http://arxiv.org/abs/gr-qc/0110100} {arXiv:gr-qc/0110100} \BibitemShut
  {NoStop}%
\bibitem [{\citenamefont {Yunes}\ \emph {et~al.}(2009)\citenamefont {Yunes},
  \citenamefont {Arun}, \citenamefont {Berti},\ and\ \citenamefont
  {Will}}]{Yunes:2009yz}%
  \BibitemOpen
  \bibfield  {author} {\bibinfo {author} {\bibfnamefont {N.}~\bibnamefont
  {Yunes}}, \bibinfo {author} {\bibfnamefont {K.}~\bibnamefont {Arun}},
  \bibinfo {author} {\bibfnamefont {E.}~\bibnamefont {Berti}}, \ and\ \bibinfo
  {author} {\bibfnamefont {C.~M.}\ \bibnamefont {Will}},\ }\href {\doibase
  10.1103/PhysRevD.80.084001} {\bibfield  {journal} {\bibinfo  {journal} {Phys.
  Rev.}\ }\textbf {\bibinfo {volume} {D80}},\ \bibinfo {pages} {084001}
  (\bibinfo {year} {2009})},\ \Eprint {http://arxiv.org/abs/0906.0313}
  {arXiv:0906.0313 [gr-qc]} \BibitemShut {NoStop}%
\bibitem [{\citenamefont {Tichy}\ \emph {et~al.}(2003)\citenamefont {Tichy},
  \citenamefont {Bruegmann}, \citenamefont {Campanelli},\ and\ \citenamefont
  {Diener}}]{Tichy:2002ec}%
  \BibitemOpen
  \bibfield  {author} {\bibinfo {author} {\bibfnamefont {W.}~\bibnamefont
  {Tichy}}, \bibinfo {author} {\bibfnamefont {B.}~\bibnamefont {Bruegmann}},
  \bibinfo {author} {\bibfnamefont {M.}~\bibnamefont {Campanelli}}, \ and\
  \bibinfo {author} {\bibfnamefont {P.}~\bibnamefont {Diener}},\ }\href
  {\doibase 10.1103/PhysRevD.67.064008} {\bibfield  {journal} {\bibinfo
  {journal} {Phys. Rev.}\ }\textbf {\bibinfo {volume} {D67}},\ \bibinfo {pages}
  {064008} (\bibinfo {year} {2003})},\ \Eprint
  {http://arxiv.org/abs/gr-qc/0207011} {arXiv:gr-qc/0207011} \BibitemShut
  {NoStop}%
\bibitem [{\citenamefont {Kelly}\ \emph {et~al.}(2007)\citenamefont {Kelly},
  \citenamefont {Tichy}, \citenamefont {Campanelli},\ and\ \citenamefont
  {Whiting}}]{Kelly:2007uc}%
  \BibitemOpen
  \bibfield  {author} {\bibinfo {author} {\bibfnamefont {B.~J.}\ \bibnamefont
  {Kelly}}, \bibinfo {author} {\bibfnamefont {W.}~\bibnamefont {Tichy}},
  \bibinfo {author} {\bibfnamefont {M.}~\bibnamefont {Campanelli}}, \ and\
  \bibinfo {author} {\bibfnamefont {B.~F.}\ \bibnamefont {Whiting}},\ }\href
  {\doibase 10.1103/PhysRevD.76.024008} {\bibfield  {journal} {\bibinfo
  {journal} {Phys. Rev.}\ }\textbf {\bibinfo {volume} {D76}},\ \bibinfo {pages}
  {024008} (\bibinfo {year} {2007})},\ \Eprint {http://arxiv.org/abs/0704.0628}
  {arXiv:0704.0628 [gr-qc]} \BibitemShut {NoStop}%
\bibitem [{\citenamefont {Kelly}\ \emph {et~al.}(2010)\citenamefont {Kelly},
  \citenamefont {Tichy}, \citenamefont {Zlochower}, \citenamefont
  {Campanelli},\ and\ \citenamefont {Whiting}}]{Kelly:2009js}%
  \BibitemOpen
  \bibfield  {author} {\bibinfo {author} {\bibfnamefont {B.~J.}\ \bibnamefont
  {Kelly}}, \bibinfo {author} {\bibfnamefont {W.}~\bibnamefont {Tichy}},
  \bibinfo {author} {\bibfnamefont {Y.}~\bibnamefont {Zlochower}}, \bibinfo
  {author} {\bibfnamefont {M.}~\bibnamefont {Campanelli}}, \ and\ \bibinfo
  {author} {\bibfnamefont {B.~F.}\ \bibnamefont {Whiting}},\ }\href {\doibase
  10.1088/0264-9381/27/11/114005} {\bibfield  {journal} {\bibinfo  {journal}
  {Class. Quantum Grav.}\ }\textbf {\bibinfo {volume} {27}},\ \bibinfo {pages}
  {114005} (\bibinfo {year} {2010})},\ \Eprint {http://arxiv.org/abs/0912.5311}
  {arXiv:0912.5311 [gr-qc]} \BibitemShut {NoStop}%
\bibitem [{\citenamefont {Mundim}\ \emph {et~al.}(2011)\citenamefont {Mundim},
  \citenamefont {Kelly}, \citenamefont {Zlochower}, \citenamefont {Nakano},\
  and\ \citenamefont {Campanelli}}]{Mundim:2010hu}%
  \BibitemOpen
  \bibfield  {author} {\bibinfo {author} {\bibfnamefont {B.~C.}\ \bibnamefont
  {Mundim}}, \bibinfo {author} {\bibfnamefont {B.~J.}\ \bibnamefont {Kelly}},
  \bibinfo {author} {\bibfnamefont {Y.}~\bibnamefont {Zlochower}}, \bibinfo
  {author} {\bibfnamefont {H.}~\bibnamefont {Nakano}}, \ and\ \bibinfo {author}
  {\bibfnamefont {M.}~\bibnamefont {Campanelli}},\ }\href {\doibase
  10.1088/0264-9381/28/13/134003} {\bibfield  {journal} {\bibinfo  {journal}
  {Class. Quantum Grav.}\ }\textbf {\bibinfo {volume} {28}},\ \bibinfo {pages}
  {134003} (\bibinfo {year} {2011})},\ \Eprint {http://arxiv.org/abs/1012.0886}
  {arXiv:1012.0886 [gr-qc]} \BibitemShut {NoStop}%
\end{thebibliography}%

\end{document}